\documentclass[mathleft,fleqn,%
]{an}
\usepackage{graphicx}
\usepackage[varg]{txfonts}
\usepackage{mathrsfs,amssymb,amsbsy}
\usepackage{graphicx}
\usepackage{subfigure}
\usepackage{paralist}
\usepackage{color}
\usepackage{natbib}
\begin{document}

\Pagespan{1}{}
\Yearpublication{2016}%
\Yearsubmission{2016}%
\Month{0}%
\Volume{999}%
\Issue{0}%
\DOI{asna.201600000}%

\title{Constraining the Milky Way assembly history with Galactic \\
Archaeology}
\subtitle{Ludwig Biermann Award Lecture 2015}

\author{I.\ Minchev\thanks{Corresponding author:
        {iminchev@aip.de}}
}
\titlerunning{Constraining the Milky Way}
\authorrunning{Minchev et al.}
\institute{
Leibniz-Institut f\"{ur} Astrophysik Potsdam (AIP), An der Sternwarte 16, D-14482, Potsdam, Germany
}
\received{2016 Apr 16}
\accepted{2016 Jun}
\publonline{2016 Aug 04}


\abstract{%
The aim of Galactic Archaeology is to recover the evolutionary history of the Milky Way from its present day kinematical and chemical state. Because stars move away from their birth sites, the current dynamical information alone is not sufficient for this task. The chemical composition of stellar atmospheres, on the other hand, is largely preserved over the stellar lifetime and, together with accurate ages, can be used to recover the birthplaces of stars currently found at the same Galactic radius. In addition to the availability of large stellar samples with accurate 6D kinematics and chemical abundance measurements, this requires detailed modeling with both dynamical and chemical evolution taken into account. An important first step is to understand the variety of dynamical processes that can take place in the Milky Way, including the perturbative effects of both internal (bar and spiral structure) and external (infalling satellites) agents. We discuss here (1) how to constrain the Galactic bar, spiral structure, and merging satellites by their effect on the local and global disc phase-space, (2) the effect of multiple patterns on the disc dynamics, and (3) the importance of radial migration and merger perturbations for the formation of the Galactic thick disc. Finally, we discuss the construction of Milky Way chemo-dynamical models and relate to observations. 
}
\maketitle

\section{Introduction}

The goal of Galactic Archeology \citep{freeman02} is to dissect the Milky Way into its various components (discs, bulge, bar and halo) and thus to disentangle the various processes that contributed to their formation and evolution. Galactic Archaeology relies on the assumptions that (i) the dynamics of formation is locked in the phase-space structure of stellar populations and that (ii) stellar atmospheres preserve the chemical imprint of their birth cloud for most of their lifetime. Chemical elements synthesized inside stars are later injected into the interstellar medium (ISM) and incorporated into the next generations of stars. Because different elements are released into the ISM by stars of different masses and on different timescales, stellar abundance ratios are thus directly related to the star formation and gas accretion history. Because stars move away from their birthplaces (a process known as radial migration), chemical information is crucial for understanding the Galactic formation history.

The importance of this topic is manifested in the number of Galactic surveys dedicated to obtaining spectroscopic information for a large number of stars, e.g., RAVE \citep{steinmetz06}, SEGUE \citep{yanny09}, APOGEE \citep{majewski10}, HERMES \citep{freeman10}, Gaia-ESO \citep{gilmore12}, and LAMOST \citep{zhao06}. This effort will soon be complemented by more than a billion stars observed by the Gaia space mission \citep{perryman01}. Millions of these will have accurate proper motions and parallaxes, which together with existing spectroscopic data, and especially with the advent of the dedicated Gaia follow-up ground-based surveys WEAVE \citep{dalton12} and 4MOST \citep{dejong12}, will enable Galactic Archaeology as never before.  

Before we can hope to understand the past Milky Way history we need a good understanding of its present state, in particular of its disc, where the majority of baryons are concentrated. This is already not a trivial task, due to the Sun's position close to the Galactic disc midplane - we cannot simply observe the disc morphology as we do in external face-on galaxies. Therefore, mostly indirect methods have been used to constrain the Galactic bar and spiral structure. 

While in axisymmetric discs energy and angular momentum are conserved quantities and are, thus, integrals of motion \citep{bt08}, this is not true for the more realistic case of potentials including perturbations from a central bar and/or spiral arms. In the case of one periodic perturbation there is still a conserved quantity in the reference-frame rotating with the pattern -- the Jacobi integral $J=E-L\Omega_p$, where $E$ is the energy of the particle, $L$ is its angular momentum, and $\Omega_p$ is the pattern angular velocity. This is no longer the case, however, when a second perturbation with a different patterns speed is included. 

It has now been well established that the Milky Way disc contains both a bar (as in more than 50\% of external disc galaxies) and spiral structure moving at different pattern speeds, making it difficult to solve such a dynamical system analytically. Instead, different types of numerical methods are usually employed, from simple test-particle integrations, to preassembled N-body and SPH systems, to unconstrained, fully cosmological simulations of galaxy formation. All of these techniques have their strengths and weaknesses. Test particles are computationally cheap, allow for full control over the simulation parameters (such as spiral and bar amplitude, shape, orientation and pattern speed) but lack self-gravity. N-body simulations offer self-consistency but bar and especially spiral structure parameters are not easy to derive and not well controlled. Finally, in addition to being very computationally intensive, the outcomes of hydrodynamical cosmological simulations are even less predictable, with merging satellites and infalling gas making it yet harder to disentangle the disc dynamics; these are, however, much closer to reality in their complexity and a necessary ultimate step in the interpretation of observational data.

Before we consider more complex systems, we first present simple test-particle models that illustrate the effect of bar and spiral density waves. 

\subsection{Resonances in galactic discs}

Galactic discs rotate differentially with nearly flat rotation curves, i.e., constant circular velocity as a function of galactic radius. In contrast, density waves, such as a central bar and spiral structure, rotate as solid bodies. Therefore stars at different radii would experience different forcing due to the non-axisymmetric structure. Of particular interest are locations in the disc where the stars are in resonance with the perturber. The corotation resonance (CR), where stars move with the pattern, occurs when the angular rotation rate of stars equals that of the perturber. The Lindblad resonances (LRs) occur when the frequency at which a star feels the force due to a perturber coincides with the star's epicyclic frequency, $\kappa$. As one moves inward or outward from the CR circle, the relative frequency at which a star encounters the perturber increases. There are two values of $r$ for which this frequency is the same as the radial epicyclic frequency. This is where the inner and outer Lindblad resonances (ILR and OLR) are located. Quantitatively, LRs occur when the pattern speed $\Omega_{p}=\Omega\pm\kappa/m$, where $m$ is the multiplicity of the pattern\footnote{$m=2$ for a bar or a two-armed spiral structure and $m=4$ for a four-armed spiral.}. The negative sign corresponds to the ILR and the positive to the OLR. While Bertil Lindblad defined these for the case of an $m=2$ pattern (thus strictly speaking the ILR/OLR are the 2:1 resonances), for an $m=4$ pattern the ILR/OLR must be the 4:1 resonances.

\begin{figure}
\centering
\includegraphics[width=\linewidth]{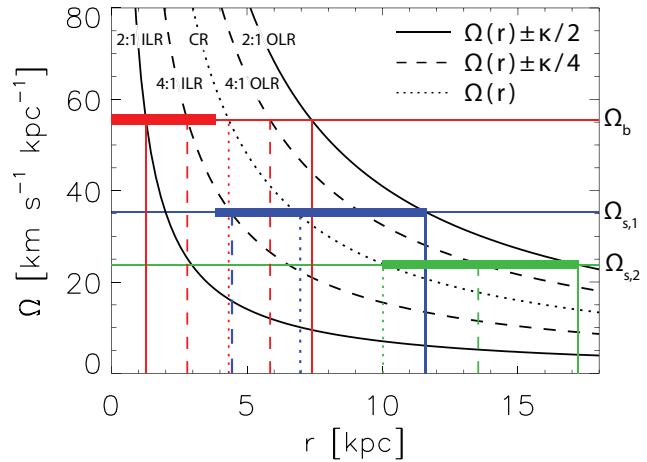}
\caption{
Resonances in a galactic disc for nearly circular orbits and a flat rotation curve. Corotation occurs along the dotted black curve and is given by $\Omega_{p}=\Omega(r)$, where $\Omega_{p}$ is the pattern speed, $\Omega(r)=v_0/r$ is the local circular frequency, and $v_0$ is the constant circular velocity. The 2:1 outer and inner Lindblad resonances (OLR and ILR) occur along the solid black curves, computed as $\Omega_{p}=\Omega(r)\pm\kappa/2$, where $\kappa$ is the local radial epicyclic frequency. The outer and inner 4:1~LRs occur along the dashed black curves, given by $\Omega_{p}=\Omega(r)\pm\kappa/4$. The red horizontal line indicates a bar pattern speed $\Omega_b=55.5$~km/s/kpc (likely for the Milky Way). An inner and outer spiral structure, moving with different pattern speeds are also shown by the blue and green lines, respectively. This is a typical situation seen in N-body simulations for multiplicity of $m=2$. The vertical red, blue, and green lines give the radial positions of each resonance for the bar and spiral structure, respectively (solid lines: 2:1; dashed lines: 4:1, dotted lines: CR). 
\label{fig:res}
}
\end{figure}

Since second order resonances, i.e., 4:1 for a two-armed spiral or bar, or 2:1 for a four-armed spiral, can also be quite important (as will be shown in \S\ref{sec:uv}), we need a convenient way to refer to them. It is somewhat confusing and unclear how the 4:1 resonances are referred to in the literature. The inner 4:1 resonance for an $m=2$ pattern is known as the {\it Ultra-harmonic resonance} (UHR). Some also describe the inner and outer 4:1 resonances as the IUHR and the OUHR, others as the inner and outer $m=4$ resonance. If the pattern multiplicity is $m=4$, then these become the ILR and OLR. To our knowledge, there is no terminology for the 2:1 resonances of an m=4 pattern. \cite{mf10} proposed to generalize the standard notation of Lindblad resonances by allowing to refer to both 2:1 and 4:1 resonances, regardless of the multiplicity of the pattern. By adopting this nomenclature, we will refer to the 2:1~ILR/OLR and the 4:1~ILR/OLR for both two-armed (or bar) and four-armed spiral structure. Naturally, other resonances can also be described in this manner, e.g., 3:1, 5:1, 6:1~ILR/OLR.  

Fig.~\ref{fig:res} illustrates the relationship between the pattern angular velocity and the radii at which resonances occur for a flat rotation curve and nearly circular orbits. The red horizontal line indicates a bar pattern speed $\Omega_b=55.5$~km/s/kpc (likely for the Milky Way). Inner and outer spiral structures moving with different pattern speeds are also shown by the blue and green lines, respectively. The actual extent of the patterns is indicated by the solid horizontal lines. This is a typical situation seen in N-body simulations for multiplicity of $m=2$ (e.g., \citealt{quillen11}) and a possible configuration for the Milky Way. The vertical red, blue, and green lines give the radial positions of each resonance for the bar, inner, and outer spiral structure, respectively (solid lines: 2:1; dashed lines: 4:1, dotted lines: CR). Note that for two (or more) non-axisymmetric patterns moving at different angular velocities there will always exist regions in the disc where resonances overlap. 

\subsection{The Milky Way bar and spirals}
\label{sec:spbar}

Due to our position in the Galactic disc, the properties of the Milky Way bar are hard to observe directly. Hence its parameters, such as orientation and pattern speed, have been inferred indirectly from observations of the inner Galaxy (e.g., \citealt{blitz91,weinberg92}). The bar has also been found to affect the local velocity distribution of stars. The way this works is as follows. If the Sun happened to lie close to a Lindblad resonance, then the local stellar velocity distribution would exhibit clumps belonging to two different orbital families, which are on nearly closed orbits in the reference frame moving with the pattern (bar or spiral structure). In the case of an OLR, a family of nearly closed orbits supporting the spiral/bar orientation exists, while a second family is misaligned with the structure outside the OLR. These are the $x_1(1)$ and $x_1(2)$ orbits, respectively (e.g., \citealt{dehnen00,fux01,minchev10}; see Fig.~\ref{fig:bar1}, left). Near the ILR the situation is similar but reversed, where the inner orbital family supports the structure but the outer one is misaligned with it.

Hipparcos \citep{perryman97} + GCS \citep{nordstrom04} data revealed more clearly a stream of old disc stars with an asymmetric drift of about 45 km/s and a radial velocity $u<0$, with $u$ and $v$ positive toward the Galactic center and in the direction of Galactic rotation, respectively. This concentration of stars in the "$u-v$ plane" is known as the Hercules stream. The numerical work of \cite{dehnen99,dehnen00,fux01,minchev10}, and more recently \cite{antoja12}, has shown that this stream can be explained as the effect of the Milky Way central bar if the Sun is placed just outside the 2:1 Outer Lindblad Resonance (OLR). Due to the inhomogeneity in age and metallicity of Hercules stars, a number of works (e.g., \citealt{famaey05, famaey07, bensby07}) have concluded that a dynamical effect, such as the influence of the bar, is a more likely explanation than a dispersed cluster. Most estimates agree on a bar orientation\footnote{See Fig.~\ref{fig:bar1}, second panel, for the definition of the bar angle.} of $\phi_b=30^\circ\pm10$ and a pattern speed of $\Omega_b/\Omega_0=1.9\pm0.1$, where $\Omega_0$ is the local standard of rest (LSR) rotation rate. A drastically different bar pattern speed has recently been suggested by the longer bar half-length measured by \cite{wegg15}, compared to previous works -- $r_b=5.0\pm0.2$ kpc, which places the CR at 5-7 kpc. In such a case the bar 2:1 OLR would lie in the range 8.5-12 kpc and, thus, would not work as an explanation for the Hercules stream. 

The Milky Way spiral structure is more poorly known than the bar. Cepheid, HI, CO and far-infrared tracers suggest that the Milky Way disc contains a four-armed tightly wound structure (see also \citealt{vallee16}), whereas \citet{drimmel01} have shown that the near-infrared observations are consistent with a dominant two-armed structure. A dominant two-armed and a weaker four-armed structure has been proposed by \citet{amaral97}. 

Similarly to the effect of the bar, the spirals can be linked to clumps in the $u-v$ plane, as first shown by \cite{qm05}. Using an orbital weighting function technique, this work showed that a two-armed spiral density wave with pattern speed placing the Sun near the 4:1 inner Lindblad resonance can account for two major clumps in the solar neighborhood's velocity distribution: the Pleiades/Hyades moving group corresponds to the one family of orbits, and the Coma Berenices moving group corresponds to another family. Similar patter speed estimate was obtained by \cite{pompeia11} and \cite{siebert12}.

\subsection{Multiple patterns in galactic discs}

Multiple patterns in N-body simulations have been known to exist since the work of \cite{sellwood85} and \cite{sellwood88}, who found that a bar can coexist with a spiral pattern moving at a much lower angular velocity. \cite{tagger87} and \cite{sygnet88} explained this as the non-linear mode coupling between the bar and the spiral wave. These findings were later confirmed by the numerical studies of \cite{masset97} and \cite{rautiainen99}. According to the theoretical work by \cite{tagger87} and \cite{sygnet88}, two patterns can couple non-linearly as they overlap over a radial range, which coincides both with the CR of the inner one and the ILR of the outer one. This coincidence of resonances results in efficient exchange of energy and angular momentum between the two patterns. The coupling between the two patterns generates beat waves (as we describe below), also found to have LRs at the interaction radii, resulting in a strong non-linear effect even at relatively small amplitudes. \cite{rautiainen99} showed that coupling between a CR and 4:1 ILR, as well as ILR and OLR is also possible in N-body simulations. Waves couple with a selection of frequencies which optimizes the coupling efficiency. Strong exchange of energy and angular momentum is then possible among the coupled waves. 

How do multiple patterns affect the dynamics of galactic discs? \cite{quillen03} considered the dynamics of stars that are affected by perturbations from both spiral arms and a central bar by constructing a one-dimensional Hamiltonian model for the strongest resonances in the epicyclic action-angle variables. Quillen pointed out that when two perturbers with different pattern speeds are present in the disc, the stellar dynamics can be stochastic, particularly near resonances associated with one of the patterns. Similar findings were presented more recently by \cite{jalali08}. All these results are not surprising since it has already been shown by \cite{chirikov79} that in the case of resonance overlap the last KAM surface between the two resonances is destroyed, resulting in stochastic behavior. It is therefore expected that resonance overlap could give rise to both velocity dispersion increase and radial migration. Both of these possibilities are explored in \S\ref{sec:heat} and \S\ref{sec:mig}, respectively.

A comprehensive discussion of the Galactic bar and spiral parameters can be found in the recent review by \cite{bland-hawthorn16}.

\subsection{Radial migration}

The power of Galactic Archaeology has been threatened both by observational and theoretical results, showing that stars most probably move away from their birthplaces, i.e, migrate radially. Observational signatures of radial migration (or mixing) have been reported in the literature since the 1970's, with the pioneering works by \cite{grenon72, grenon89}. Grenon identified an old population of \emph {super-metal-rich stars} (hereafter SMR), presently at the Solar vicinity, but with kinematics and abundance properties indicative of an origin in the inner Galactic disc. SMR stars show metallicities which exceed the present day ISM and those of young stars at the solar vicinity. The metallicity of the solar vicinity, however, is not expected to increase much since the Sun's formation, or in the last $\sim$4~Gyr, due to the rather inefficient star formation rate (SFR) at the solar radius during this period (e.g., \citealt{chiappini03,asplund09}). Hence, pure chemical evolution models for the Milky Way thin disc cannot explain stars more metal rich than $\sim$0.2~dex and the effect of radial migration needs to be considered.

N-body simulations have also long shown that radial migration is unavoidable. \cite{raboud98} studied numerical simulations aimed at explaining the results by \cite{grenon89} of a mean positive $u$-motion, which the authors interpreted as metal-rich stars from the inner galaxy, wandering in the solar neighborhood. However, \cite{raboud98} interpreted their findings as stars on hot bar orbits, not recognizing that permanent changes to the stellar angular momenta are possible. It was not until the work by \cite{sellwood02} that radial migration was established as an important process affecting the entire disc, where stars shift guiding radii due to interaction with transient spiral structure. A number of works using N-body and N-body/SPH simulations (e.g., \citealt{roskar08a, grand12,minchev11a}) have confirm that migration always takes place in numerical galactic discs. 

A different radial migration mechanism was proposed by \cite{mf10}, who considered the simultaneous propagation of a bar and a {\it long-lived} spiral density wave. In such a setup angular momentum redistribution arises from the overlap of resonances associated with different modes or from the stellar mass transiency as perturbers with different pattern speeds interfere constructively (see \citealt{comparetta12}), but not from the growth and decay of transient modes. This work, along with studies of diffusion coefficients in barred discs \citep{brunetti11,shevchenko11}, predicts a variation in migration efficiency with time and disc radius, establishing that the dynamical influence of the bar plays an integral part of Milky Way disc modeling. Aside from internal structure, perturbations due to minor mergers have also been shown to be effective at mixing the outer discs \citep{quillen09, bird12}, but also can, indirectly, affect the entire disc by inducing (reinforcing) spiral and bar instabilities (e.g., \citealt{purcell11}). Considering the established presence of a central bar, spiral structure and evidence for merger activity in the Milky Way, it is clear that all of the above mentioned radial migration mechanisms would have an effect on the Galactic disc.

Stars found today at a given small Galactic disc region (e.g., the solar neighborhood) can have a range of birth radii but are mostly indistinguishable in their kinematics from locally born stars. Therefore, chemical abundances need to be invoked in identifying migrators. Probably one of the best ways of quantifying radial migration in the Milky Way is using the technique of chemical tagging \citep{bland-hawthorn10}, where stars born in the same cluster (now dispersed) are expected to appear as a clump in the multi-dimensional chemical space. This is one of the main objectives of the ongoing GALAH survey \citep{freeman10}.

\begin{figure*}
\centering
\includegraphics[width=\linewidth]{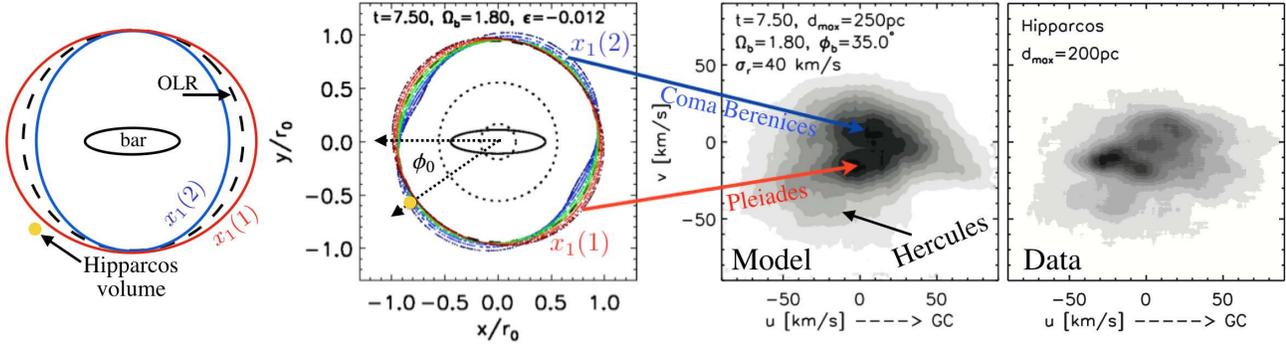}
\caption{
The left panel Illustrates the $x_1(1)$ and $x_1(2)$ orbits and their orientation with respect to the bar. The dashed circle depicts the OLR position (about 800 pc inside the solar circle) such that the Hercules moving group can be explained by the bar. 
The second panel shows the results of a test-particle simulation from \cite{minchev10} at time $t=7.5$ (in units of solar rotations). The axes are in units of the the solar radius, $r_0$. The dotted circle shows the bar's CR. Only stars initially on circular orbits close to the OLR are shown, with those inside/outside the OLR colored in blue/red. The bar is fully grown in four bar rotations, which corresponds to $t\approx2.2$ in the units used here. Note that the $x_1(1)$ and $x_1(2)$ orbits orientation is not as in the left panel but offset by about $30^\circ$ in the clockwise direction. The approximate Hipparcos coverage is shown by the yellow-filled small circle. 
The third panel shows the simulated $u-v$ plane at the depicted time. The radial velocity dispersion is $sigma_r=40$~km/s, in order to populate the Hercule stream as well. The pattern speed is fixed at $\Omega_b/\Omega_0=1.8$, the maximum sample depth is $d_{max}=250$ pc, and the bar orientation is $\phi_0=35^\circ$. Shaded contours show the particle number density. 
The two orbital families associated with the $x_1(1)$ (red) and $x_1(2)$ (blue) orbits precess at different rates and provide good match to the Coma Berenices and Pleiades groups at this particular time. 
Hipparcos stellar velocity distribution with the Sun's motion subtracted (values from \citealt{db98}).
}
\label{fig:bar1}      
\end{figure*}

\subsection{The Galactic thick disc}

The formation of galactic thick discs has been an important topic ever since their discovery in external galaxies \citep{burstein79, tsikoudi79} and in the Milky Way \citep{gilmore83}. The large uncertainties in important observational constraints in the Milky Way, such as the age-velocity-metallicity relation, abundance gradients and their evolution, have led to different scenarios to be proposed for the formation of the Galactic thick disc. 

One possibility is that the stars comprising thick discs are born thick at high redshift from internal gravitational instabilities in gas-rich, turbulent, clumpy discs \citep{bournaud09, forbes12} or in the turbulent phase associated with numerous gas-rich mergers \citep{brook04,brook05}. They could also have been created through accretion of galaxy satellites \citep{meza05, abadi03}, where thick disc stars then have an extragalactic origin. 

Another possibility is that thick discs are created through the heating of preexisting thin discs by minor mergers \citep{villalobos08, dimatteo11}. Evidence for merger encounters can be found in the phase-space structure of Milky Way disc stars (e.g., \citealt{minchev09, gomez12a, gomez13}).

Several works (e.g., \citealt{schonrich09b, loebman11}) have also proposed that radial migration can give rise to thick disc formation by bringing out high-velocity-dispersion stellar populations from the inner disc and the bulge. More detailed dynamical studies \citep{minchev12b, martig14b, vera-ciro14} have more recently shown that migration does not contribute to any significant level to disc thickening, but on the opposite, it suppresses flaring when external perturbations are included \citep{mcm14,grand16}. This is discussed further in \S\ref{sec:mig_thick}.

Finally, a recently proposed thick disc formation mechanism is the superposition of coeval flaring disc subpopulation \citep{minchev15}. Disc flaring can result from a number of different sources, the most effective most likely being the perturbative effect of mergers on the host disc. In an inside-out formation scenario, the old populations dominate in the inner disc and younger in the outer disc, thus the flaring disappears when the total stellar population is taken into account. By making the distinction between a thick disc defined as the [$\alpha$/Fe]-high, old stellar population in the Milky Way and a geometrically defined thick discs in observations of edge-on galaxies, this new view of the formation of thick discs resolves a number of apparent contradictions. More on this can be found in \S\ref{sec:thick}.

\section{Constraining the Galactic bar}
\label{sec:bar}
 
\begin{figure*}
\centering
\includegraphics[width=\linewidth]{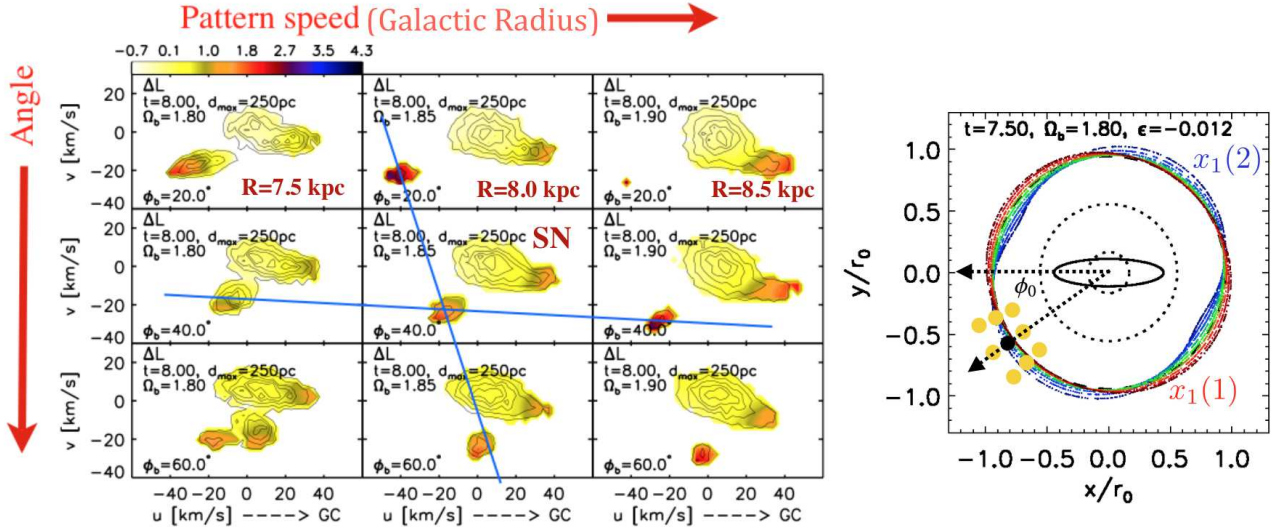}
\caption{
{\bf Left}: Variation in the $u-v$ plane with bar pattern speed and orientation. Contours show particle number density, while the color levels represent the change in angular momentum $\Delta L$ for a given location on the $u-v$ plane. Color bar values can be converted to (km/s pc) by multiplying by $100 v_0$, with $v_0$ the rotation curve. Different rows show different bar major axis orientations of $\phi_0=20^\circ, 30^\circ$ and $60^\circ$ with respect to the Sun Galactocentric line. Different columns show changes in bar angular velocity $\Omega_b/\Omega_0=1.8,1.85,$ and $1.9$, where the bar's OLR is at $\Omega_b/\Omega_0=1.7$. These correspond approximately to neighborhoods at Galactic radii $r=7.5, 8.0$ and $8.5$. The blue lines indicated the systematic shifts of clumps in the $u-v$ plane with change in bar angel and distance from the Galactic center.  
{\bf Right}: As the second panel in Fig.~\ref{fig:bar1}, but with the nine neighborhood locations corresponding to the $u-v$ plots shown on the left overlaid as yellow-filled circles (black for the solar neighborhood). Adapted from \cite{minchev10}.
\label{fig:bar2}
}
\end{figure*}

\subsection{The effect of a recently formed bar}
\label{sec:bar1}

While earlier work has shown the bar to affect the $u-v$ plane at higher velocities, some of the low-velocity moving groups in the solar vicinity have been explained only with the effect of spiral density waves, such as the splitting of the Pleiades/Hyades and Coma Berenices moving groups (e.g., \citealt{qm05}). The reason the bar's OLR should not affect stars with kinematics cooler than the Hercules stream is that the solar radius lies about 0.8 kpc outside the OLR (dashed circle), thus, stars on near circular orbits at the OLR cannot reach the solar neighborhood. This situation can be seen in the leftmost panel of Fig.~\ref{bar1}, where the Hipparcos volume is shown by the yellow-filled small circle. Hotter stars near the OLR, however, can appear close to the Sun on their apocenters. 

Studying the effect of the bar, we explored a different scenario in \cite{minchev10}, where the time evolution of the $u-v$ plane was examined just after the bar formation. These test-particle simulations were performed in a Milky Way-like potential to which a perturbation due to a central bars was added, similar to \cite{dehnen00}. The background axisymmetric potential due to the disc and halo has the form
\begin{equation}
\Phi_0(r)=v_0^2\log(r),
\label{eq:axi}
\end{equation}
corresponding to a flat rotation curve. We model the nonaxisymmetric potential perturbation due to the Galactic bar as a pure quadrupole
\begin{equation}
\Phi_{\rm b} = A_{\rm b}(\epsilon_{\rm b}) \cos[2(\phi-\Omega_{\rm b}
t)]\times\left\{
\begin{array}{cclcr}
         \left(r_{\rm b}\over r\right)^3  &,&  r&\ge & r_{\rm b}    \\ 
       2-\left(r\over r_{\rm b}\right)^3  &,&  r&\le & r_{\rm b}
\end{array}
\right.
\label{eq:bar}
\end{equation}
Here $A_{\rm b}(\epsilon_{\rm b})$ is the bar's gravitational potential amplitude, identical to the same name parameter used by \cite{dehnen00}; the strength is specified by $\epsilon_{\rm b}=-\alpha$ from the same paper. The bar length is $r_{\rm b}=0.8r_{\rm cr}$ with $r_{\rm cr}$ the bar corotation radius. The pattern speed, $\Omega_{\rm b}$ is kept constant. The bar amplitude $\epsilon$ is initially zero, grows linearly with time at $0<t<t_1$ and transitions smoothly to a constant value after $t=t_1=4$ bar rotations. This insures a smooth transition from the axisymmetric to the perturbed state.

Previous work has used as a measure of bar strength the parameter $Q_T$ \citep{combes81}. This is the ratio of the maximum tangential force to the azimuthally averaged radial force at a given radius. From eq.~\ref{eq:bar} this definition yields $Q_T=2A_b/v_c^2$. We examine bar amplitudes in the range $0.1<Q_T<0.4$ as expected from observations of various galaxies and from N-body simulations \citep{combes81}. This corresponds to $0.013<|\epsilon_b|<0.05$. 

In our units the solar neighborhood radius is $r_0=1$; the circular speed is $v_0=1$ everywhere since the rotational curve is flat. To convert to real units we use a LSR tangential velocity of 240 km/s, and Galactocentric distance of 8 kpc. The 2:1 OLR with the bar is achieved when $\Omega_{\rm b}/\Omega_0=1+\kappa/2\approx1.7$, where $\kappa$ is the epicyclic
frequency. For a flat rotation curve $\kappa=\sqrt{2}\Omega_0$, where $\Omega_0$ is the angular velocity of the LSR.

Interestingly, we found that a steady state bar induced transient features at low velocities in the simulated solar neighborhood velocity distribution due to the initial response of the disc to the bar formation. We associate these velocity streams with two quasi-periodic orbital families librating around the stable $x_1(1)$ and $x_1(2)$ orbits near the bar's OLR (Fig.~\ref{fig:bar1}, left two panels). In a reference frame moving with the bar these, otherwise stationary, orbits precess on a timescale dependent on the strength of the bar. The effect of this precession can be seen in Fig.~\ref{fig:bar1}, where the left panel illustrates the $x_1(1)$ and $x_1(2)$ orbits and their orientation with respect to the bar. The second panel shows the results of a test-particle simulation from \cite{minchev10} at time $t=7.5$ (in units of solar rotations). Only stars initially on circular orbits close to the OLR are shown, with those inside/outside the OLR colored in blue/red. Note that, because of the precession, at this time the $x_1(1)$ and $x_1(2)$ orbits orientation is not as in the left panel but offset by about $30^\circ$ in the clockwise direction. This behavior allows the two (kinematically cold) orbital families to reach the solar neighborhood and manifest themselves as clumps in the $u-v$ plane moving away from ($x_1(2)$), and toward ($x_1(1)$) the Galactic center. 

The nine panels on the left in Fig.~\ref{fig:bar2} show the simulated $u-v$ plane for nine neighborhoods at different Galactic disc positions, as indicated by the yellow circles in the x-y plot on the right. The solar neighborhood volume is shown as a black circle. For this particular simulation, only test-particles on nearly circular orbits were subjected to the potential in order to avoid the sea of hot orbits, which we already know give rise to Hercules stream-like feature (see Fig.~\ref{fig:bar1} third panel). 

\begin{figure*}
\centering
\includegraphics[width=\linewidth]{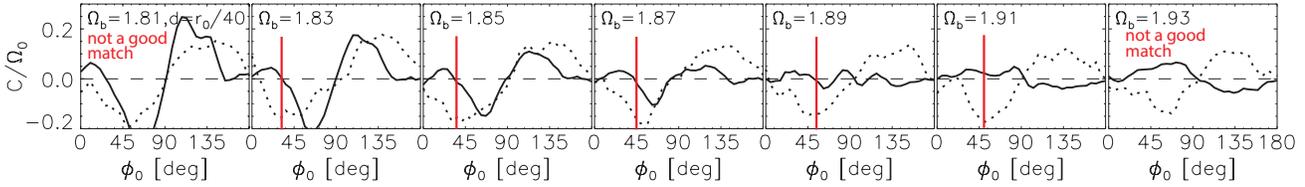}
\caption{
Each panel shows the variation of the Oort constant $C$ with bar angle $\phi_0$, for a simulation with a particular bar pattern speed, $\Omega_{\rm b}$, and a mean heliocentric distance, $\overline{d}=r_0/40$, or 200 pc for a Galactocentic radius $r_0=8$~kpc. Solid and dotted lines correspond to stellar populations with cold and hot kinematics values, respectively. Panels from left to right show an increasing $\Omega_{\rm b}$ in units of $\Omega_0$. Note that the OLR is at $\Omega_{\rm OLR}\approx1.7$. Good matches to the observed trend in $C$ (vanishing value for the cold disc and
a large negative for the hot one) are achieved for $20^\circ\leqslant\phi_0\leqslant45^\circ$ and $1.83\leqslant\Omega_{\rm b}/\Omega_0\leqslant1.91$. These are indicated by the red vertical lines. Figure is a modified version of Fig.~2 by \cite{mnq07}.
}
\label{fig:c}      
\end{figure*}

For a given time, the tangential velocity $v$ of resonant features in the $u-v$ plane is set by the bar pattern speed and the radial velocity $u$, is set by the bar's orientation. These variations are illustrated by the blue lines in the figure. Thus, assuming some features at low velocities in the Hipparcos velocity distribution are of resonant origin, we can match stream positions in the $u-v$ plane and estimate the bar pattern speed $\Omega_b$ and the orientation of its major axis with respect to the solar Galactocentric line, $\phi_0$. In addition, structure varies with time due to the libration of the quasi-periodic orbits around the fixed points, which allows us to constrain the bar formation time.

Depending on the bar parameters and time since its formation, this model is consistent with the Pleiades and Coma Berenices, or Pleiades and Sirius moving groups seen in the Hipparcos stellar velocity distribution, when the Milky Way bar angle is $30^\circ\la\phi_0\la45^\circ$ and its pattern speed is $\Omega_b/\Omega_0=1.82\pm0.07$. Since the process is recurrent,  a good match could be achieved about every six LSR rotations. However, to be consistent with the fraction of stars in the Pleiades, we estimated that the Milky Way bar formed $\sim2$ Gyr ago. This model argues against a common dynamical origin for the Hyades and Pleiades moving groups.

We the advent of Gaia accurate proper motions and parallaxes, combined with high precision line-of-sight velocities from APOGEE and RAVE (already after Gaia DR1 and DR2) and ultimately by the 4MOST survey (scheduled to start operation in 2021), we will be able to test these predictions by trying to reproduce the variation of structure in the $u-v$ plane at different positions in the disc (as in the left half of Fig.~\ref{fig:bar2}).

\subsection{Relation to the Oort C constant}
\label{sec:bar2}

As discussed above, previous work has related the Galactic bar to structure in the local stellar velocity distribution. 

Another indirect way to constrain the bar was presented by \cite{mnq07}, where we showed that the bar also influences the spatial gradients of the velocity vector via the Oort constants. 

We can linearize the local velocity field about the LSR and write the mean radial velocity $\overline{v}_d$ and longitudinal proper motion $\overline{\mu}_l$ as functions of the Galactic longitude $l$ as
\begin{eqnarray}
\label{eq:vel}
{\overline{v}_d\over \overline{d}} &=& K + A\sin(2l) + C\cos(2l)   \\
\overline{\mu}_l &=& B + A\cos(2l) - C\sin(2l)\nonumber
\end{eqnarray}
where $\overline{d}$ is the average heliocentric distance of stars, $A$ and $B$ are
the usual Oort constants, and $C$ and $K$ are given by
\begin{eqnarray}
\label{eq:oc}
2C  &\equiv &   -{\overline{u} \over r} + {\partial \overline{u} \over \partial r}
- {1\over r} {\partial \overline{v}_{\phi} \over \partial \phi}         \\
2K  &\equiv &   +{\overline{u} \over r} + {\partial \overline{u} \over \partial r}
+ {1\over r} {\partial \overline{v}_{\phi} \over \partial \phi}.
\end{eqnarray}
Here $r$ and $\phi$ are the usual polar coordinates and $v_\phi=v_0+v$, where $v_0$ is the circular velocity at the Solar radius, $r_0$. Considering a flat rotation curve, the derivatives of $v_\phi$ in the above equations are identical to the derivatives of $v$. $C$ describes the radial shear of the velocity field and $K$ its divergence. For an axisymmetric Galaxy we expect vanishing values for both $C$ and $K$\footnote{Note, however, that $C$ and $K$ would also be zero in the presence of nonaxisymmetric structure if the Sun happened to be located on a symmetry axis.}. Whereas $C$ could be derived from both radial velocities and proper motions, $K$ can only be measured from radial velocities, in which case accurate distances are also needed.

The study of \cite{olling03} not only measured a non-zero $C$, implying the presence of non-circular motion in the local disc, but also found that $C$ is  more negative for older and redder stars with a larger velocity dispersion, while it is roughly zero for the cold (thus young) population. More recent determination of the $C$ constant using RAVE data confirmed these results \citep{siebert11b}. This variation of $C$ with velocity dispersion/color/age is surprising as a hotter stellar population is expected to have averaged properties more nearly axisymmetric, and hence, a reduced value of $|C|$ (e.g., \citealt{mq07}). 

Assuming the Galactic bar affects the shape of the distribution function of the old stellar population in the solar neighborhood, an additional constraint on the bar can be provided by requiring that a model reproduces the observed value of the Oort constant $C$. In other words, in addition to relating the dynamical influence of the bar to the local velocity field (see \S~\ref{sec:bar1} above), $C$ provides a link to the gradients of the velocities as well. 

In an effort to reproduce these observational results, using test-particles integrations we simulated measurements of the Oort $C$ value in a gravitational potential including the Galactic bar, as in \S~\ref{sec:bar1} above. For all other parameters fixed, we defined a cold and a hot sample resulting from an initial radial velocity dispersion of 10 km/s and 40 km/s, respectively.

In Fig.~\ref{fig:c} we present our results for $C$ as a function of the bar angle, $\phi_0$ (the angle by which the Sun's azimuth lags the bar's major axis). Each panel shows a simulation with a different pattern speed and sample depth of $\overline{d}=200$~pc. Solid and dotted lines represent the results for cold and hot discs, respectively. $C$ is presented in units of $\Omega_0=v_0/r_0$. To make the discussion less cumbersome, we write $C_{\rm h}$ and $C_{\rm c}$ to refer to the values for $C$ as estimated from the hot and cold discs, respectively.  

$C_{\rm h}$ (dotted lines in Fig. \ref{fig:c}) varies with galactic azimuth as $C_{\rm h}(\phi_0)\sim\sin(2\phi_0)$ for all of the $\Omega_{\rm b}$ values considered. On the other hand, the cold disc values (solid lines) exhibit different variations, depending on the bar pattern speed or equivalently, on the ratio $r_0/r_{OLR}$. Closer to the OLR (left panels of Fig. \ref{fig:c}), $C_{\rm c}(\phi_0)$ approaches the functional behavior of $C_{\rm h}(\phi_0)$. Away from the OLR (right panels), $C_{\rm c}(\phi_0)$ is shifted by $90^\circ$ compared to $C_{\rm h}(\phi_0)$, i.e., $C_{\rm c}(\phi_0)\sim-C_{\rm h}(\phi_0)$. Moreover, the hot disc yields an increase in the amplitude of $C_{\rm h}(\phi_0)$ as the pattern speed nears the OLR. Whereas the same trend is apparent for the cold disc, the gradient of the amplitude of $C_{\rm c}(\phi_0)$ is much larger. This is consistent with our expectation that the cold disc is affected more by the bar, especially near the OLR. While close to the OLR $|C_{\rm h}(\phi_0)|<|C_{\rm c}(\phi_0)|$, we observe the opposite behavior away from it. This could be explained by the results of \cite{muhlbauer03}, where it was found that high velocity dispersion stars tend to shift the ``effective resonance" radially outwards.

The red vertical lines in Fig. \ref{fig:c} indicate possible solar azimuths where the observed Oort $C$ trends are well matched. By comparing measurements of $C$ with our simulations we constrained the pattern speed as $\Omega_{\rm b}/\Omega_0=1.87\pm0.04$, where $\Omega_0$ is the local circular frequency, and found the bar angle to lie in the range $20^\circ\leqslant\phi_0\leqslant45^\circ$.

\section{Constraining the Milky Way spiral structure}
\label{sec:sp}

\begin{figure}
\centering
\includegraphics[width=6. cm]{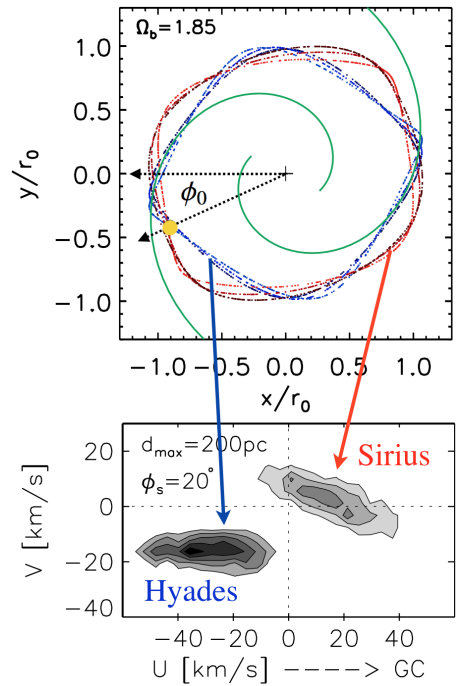}
\caption{
{\bf Top panel:} The effect of a two-armed spiral structure on orbits near the 4:1 ILR (or IUHR). Note the splitting into two families of closed orbits in the frame moving with the spiral pattern. $\phi_0$ indicates the orientation of spiral structure with respect to the Sun's azimuth, given by the angle between the dotted Galactocentic arrows. For a Sun orientation at $20^\circ$ with respect to a concave arm, both orbital families can enter the solar neighborhood (yellow filled circle). {\bf Bottom panel:} the effect on the $u-v$ plane for the configuration shown in the top panel. The clumps at $(u,v)\approx(-35, -17)$ km/s and $(U,V)\approx(10,0)$ km/s are good matches to the Hyades and Sirius moving groups, respectively. Test-particle simulation used is from \cite{mq08}. This figure is similar to Fig.~13 by \cite{pompeia11}.
}
\label{fig:sp}      
\end{figure}

\subsection{Constraints using the U-V plane}
\label{sec:uv}

Using a spiral pattern speed of $\sim20$~km/s/kpc (assuming LSR rotation of 28 km/s/kpc), \cite{qm05} used an orbital weighting function technique to show that a two-armed spiral density wave can split the solar neighborhood's velocity distribution into two major clumps. A different approach but with a similar result was presented by \cite{pompeia11} and discussed below. In that work a test-particle simulation by \cite{mq08} was employed, in which initially a disc was populated by stars on circular orbits and then numerically integrated in the axisymmetric potential from eq.~\ref{eq:axi}, but including a spiral, instead of a bar perturbation term. 

The spiral potential is given by
\begin{equation}
\Phi_s(r,\phi,t) = \epsilon_s \cos[\alpha \ln{r\over r_0}-m(\phi-\Omega_s t)],
\label{eq:sp}
\end{equation}
where $\epsilon_s$ is the spiral strength, related to the amplitude of the mass surface
density of spirals, $\Sigma_s$, as 
\begin{equation}
\label{eq:sp3}
\epsilon_s \approx - 2 \pi G  \Sigma_s r_0 / (\alpha v_c^2),
\end{equation}
as shown in \cite{bt08}. The parameter $\alpha$ is related to the pitch angle of the spirals, $p$, by $\alpha = m \cot(p)$. The azimuthal wavenumber $m$ is an integer corresponding to the number of arms. We consider both two-armed and four-armed spiral structure with $\alpha=-4$ and $-8$, respectively, where the negative sign corresponds to trailing spirals. \cite{elmegreen98} found that grand-design spirals have arm-interarm contrasts of 1.5-6, corresponding to a fractional amplitude of $0.2<\Sigma_s/\Sigma<0.7$, which is in agreement with \cite{rix95} who estimated $0.15<(\Sigma_s/\Sigma)<0.6$. 

For a maximum exponential disc the peak circular speed in the disc has the value $v_c \simeq 0.622 \sqrt{GM_d / r_d}$ at $r\simeq 2.15r_d$, where $M_d$ is the disc mass inclosed by $r_d$. The surface density of the disc at radius $r$ is $\Sigma(r) = ({M_d / 2\pi r_d^2}) e^{-r/r_d}$. Eliminating $M_d$ from these expressions, we find at $r_0$, $v_c^2 \simeq 0.39 \times 2\pi G r_d\Sigma_0 e^{r_0/r_d}$. Substituting this expression for $v_c$ in eq.~\ref{eq:sp3}, the relation between the relative potential and the relative overdensity becomes
\begin{equation}
\label{eq:spden}
\epsilon_s \approx - {\Sigma_s\over \Sigma_0} {r_0 \over r_d} {e^{-r_0\over r_d} \\
\over 0.39 \alpha}.
\end{equation}

The perturbation is grown from zero to its maximum strength in four rotation periods at $r_0$. In order to improve statistics, positions and velocities are time averaged for 10 spiral periods. We distribute particles (stars) between in inner and outer galactic radii $(r_{in}, r_{out})=(0.3r_0,2.0r_0)$. New particles are added until the final number of outputs is $2.5\times10^6$. In addition, the two-fold symmetry of our model galaxy is used to double this number. More details about these simulations can be found in \cite{mq08}. 

The top panel of Fig.~\ref{fig:sp} shows the effect of a two-armed spiral structure on orbits near the 4:1 ILR\footnote{Also known as the inner ultra-harmonic resonance or IUHR.}. This 4:1 resonance gives rise to square orbits, in the frame moving with the spiral pattern, even though the imposed spiral wave is two-armed. Similarly to the case of the bar (see \S\ref{sec:bar}), two families of closed orbits are excited by the resonance, where one supports the spiral structure (blue particles) and the other one is misaligned with it (red particles). $\phi_0$ indicates the orientation of spiral structure with respect to the Sun's azimuth, given by the angle between the dotted Galactocentic arrows. For a Sun orientation at $20^\circ$ with respect to a concave arm, both orbital families can enter the solar neighborhood (yellow circle). The Galactocentric axes are in units of $r_0$, the Galactocentric radius of the Sun. 

The bottom panel of Fig.~\ref{fig:sp} shows the effect on the $u-v$ plane for the configuration shown in the top panel. To match spatially the Hipparcos stellar sample, only particles in a 200 pc circle around the fiducial Sun are selected. Each orbital family gives rise to a stream (or moving group) in velocity space. The dense clump at $(u,v)\approx(-35, -17)$ km/s can be associated with the Hyades and the shallow one at $(u,v)\approx(10,0)$ km/s with the Sirius moving groups. The contour levels correspond to 0.2, 0.31, 0.43, 0.55, 0.67 and 0.8 of the maximum value at the centre of the Hyades clump. 

This model constrains the spiral pattern speed to $\Omega_s/\Omega_0=0.65$ and orientation with respect to the Sun, $\phi_0=20^\circ$ to an uncertainty of less than 5\%.

\begin{figure}
\includegraphics[width=\linewidth]{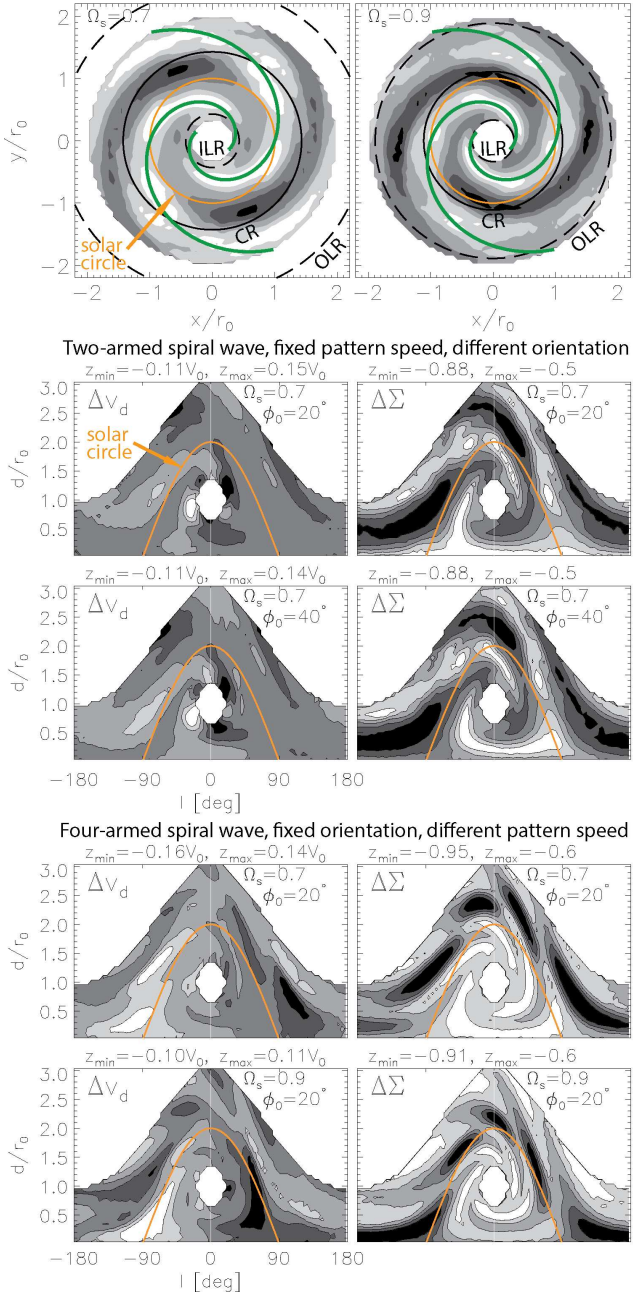}
\caption{
{\bf Top row:} Test-particle simulation of a galactic disc perturbed by a two-armed spiral density wave. Lighter colors correspond to higher stellar density. Distances are in units of the solar distance from the Galactic center, $r_0$. As the pattern speed is increased the resonances are shifted inwards bringing the CR closer to the solar circle.
{\bf Second and third rows:} Maps of the residual mean line-of-sight velocity, $\Delta V_d$, and residual stellar density, $\Delta \Sigma$, for the disc in the top-left panel, plotted versus Galactic longitude $l$ (x-axis), and heliocentric distance $d/r_0$ (y-axis). The orange curve is the projection of the solar circle. The solar phase angle is $\phi_0=20^\circ$ in the second row and $\phi_0=40^\circ$ in the third row. Minimum (black) and maximum (white) contour values are displayed on top of each panel.
{\bf Fourth and fifth rows:} Same as above, but for a four-armed structure with a phase angle of $20^\circ$ and different pattern speeds, as shown. Figure adapted from \cite{mq08}.
}
\label{fig:pencil}
\end{figure}

\subsection{The effect of spiral structure on the global disc phase space}

Another way to obtain constraints on the Milky Way spiral structure is through stellar samples covering large parts of the Galactic disc. 
To facilitate the interpretation of the huge amounts of data expected from Gaia and spectroscopic follow-up surveys, studies of how spiral structure affects the global disc phase space are needed. One such attempt was made by \cite{mq08}, where test-particle simulations of a galactic disc perturbed by a steady-state spiral density wave used to relate structure in velocity and morphology to the spiral parameters. 

In the top row of Fig~\ref{fig:pencil} we present stellar number density contour plots for two such simulations with different spiral pattern speeds, as indicated. The background axisymmetric disc is subtracted to emphasize the spiral structure. The quantity plotted is $\Delta \Sigma=(\Sigma-\Sigma_{axi})/\Sigma_{axi}$, where $\Sigma$ and $\Sigma_{axi}$ are the perturbed and axisymmetric stellar number densities.
The resonances get closer together for the faster spiral on the right. Distances are in units of the solar radius, $r_0$. Darker colors correspond to lower density. The inner $0.3r_0$ disc is not plotted since we do not model the Galactic center. Note the disruption of the spirals near the 2:1 LRs (dashed circles). 

The second and third rows of Fig~\ref{fig:pencil} show maps of the residual mean line-of-sight velocity, $\Delta V_d$, and the residual stellar density, $\Delta \Sigma$, for the disc in the top-left panel. Here, however, these quantities are plotted versus Galactic longitude $l$ (x-axis), and heliocentric distance $d/r_0$ (y-axis). The orange curve is the projection of the solar circle. The solar phase angle is $\phi_0=20^\circ$ in the second row and $\phi_0=40^\circ$ in the third row. The minimum (black) and maximum (white) contour values are displayed on top of each panel. Well defined structure is present for both $\Delta V_d$ and $\Delta \Sigma$ and that is seen to change with the change in solar orientation with respect to the spirals. 

The fourth and fifth rows of Fig~\ref{fig:pencil} are similar to the two above, but show the results for a four-armed structure with a phase angle of $20^\circ$ and a pattern speed $\Omega_s=0.7\Omega_0$ (fourth row) and $\Omega_s=0.9\Omega_0$ (fifth row). Thus, the variation in structure seen here is due to the different pattern speeds used.

Fig~\ref{fig:pencil} indicates that (1) the solar orientation with respect to the spiral, (2) the spiral pattern speed and (3) the number of spiral arms can be associated with structure in these observationally motivated maps.

\cite{mq08} found that the axisymmetric potential needs to be known to $\sim10\%$, line-of-sight velocities to $\sim20$ km/s, and distance uncertainties need to be less than $\sim30\%$, in order to be able to constrain spiral structure. The mean line-of-sight velocity and the velocity dispersion are affected by up to $\sim35$ km/s which is well within the detectable limit of even low-resolution spectroscopic surveys.

\begin{figure*}
\centering
\includegraphics[width=\linewidth]{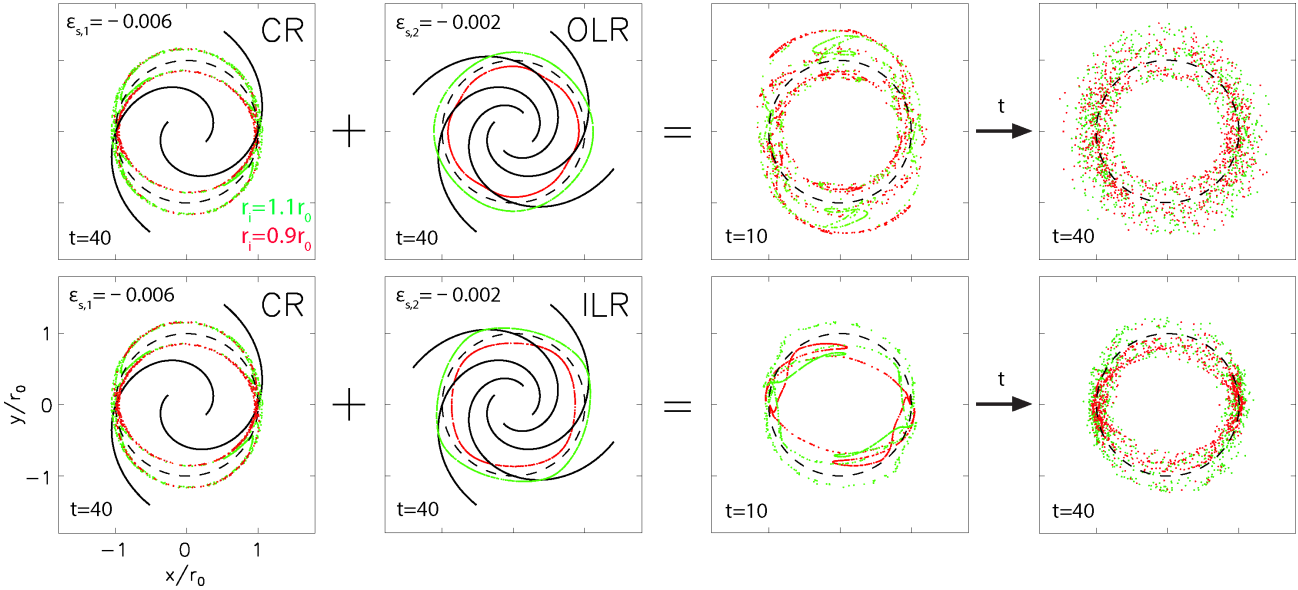}
\caption{ 
The effect on stellar orbits of two spiral perturbation with different pattern speeds acting together. {\bf First column:} The red/green particles initially start on rings just inside/outside the CR (dashed black circle) of a two-armed spiral wave. Distances are in units of the solar radius, $r_0$. The top and bottom plots are identical. {\bf Second column:} Same initial conditions, but near the 4:1 OLR (top) or 4:1 ILR (bottom) of a four-armed wave. {\bf Third column:} Same initial conditions, but particles are perturbed by both spiral waves shown at t=10 rotation periods, or about 2.5 Gyr. Note that a secondary wave of only 1/3 the strength of the first one is enough to disrupt the horseshoe orbits near the CR. {\bf Fourth column:} Same setup as in the third column but at t=40, or about 10 Gyr. Simulations from \cite{mq06}.
\label{fig:mq06}
}
\end{figure*}

A survey of stars close to the disc midplane and covering large area is need to apply this method of constraining the spirals. Such large-scale Galactic surveys have now become a reality, most notably, the APOGEE-1 near-infrared SDSS project, which covers Galactocentric azimuth of about $60^\circ$ in the radial range $3<r<14$~kpc for some $1.5\times10^5$ stars. The surveyed disc area will soon increase with the forthcoming APOGEE-2 (north and south) and the addition of $\sim3\times10^5$ more stars. In the very near future  Gaia and follow-up surveys 4MOST and WEAVE will increase this number to millions of stars with accurate proper motions, parallaxes, radial velocities, photometry, and chemical composition.

\section{The importance of multiple patterns in galactic discs}

So far we have only considered simple test-particle models, but which allow a high degree of control and the means of cheaply sweeping parameter space. Only the effect of a bar or that of a spiral perturbation was used for the models discussed in \S~\ref{sec:bar} and \S~\ref{sec:sp}. Using test-particle simulations, we could isolate, understand, and quantify better the effect of an individual perturber, i.e., the bar or a spiral wave. It is well known, however, that the Milky Way disc, as well as more than 50\% of disc galaxies in general, harbor both types of non-axisymmetric components. Moreover, both analyses of N-body simulations (e.g., \citealt{tagger87, sygnet88}) and expanding galaxy images in Fourier components (e.g, \citealt{elmegreen92,rix93}) have shown the existence of multiple spiral density waves, which can propagate simultaneously in galaxy discs.

Complicating the disc dynamics by the inclusion of multiple patterns and studying their combined effect on the kinematical heating of, and redistribution of angular momentum (radial migration) in, galactic discs is the topic of the following \S\ref{sec:heat} and \S\ref{sec:mig}, respectively. In \S\ref{sec:longevity} we show that, indeed, N-body SPH simulations support the idea that spiral patterns can be long-lived features, justifying the assumptions we have been making so far.

\subsection{Stochastic heating from multiple spiral waves}
\label{sec:heat}

A new mechanism for increasing the stellar velocity dispersion with time in galactic discs (known as the age-velocity relation) was described by \cite{mq06}, where we studied the effect of two spiral density waves propagating with different pattern speeds. The simulations performed were similar to the ones described in \S\ref{sec:sp}, however, two perturbation terms in the form of eq.~\ref{eq:sp} were added to the axisymmetric potential (eq.~\ref{eq:axi}) and evolved in time. While the two spiral perturbations were described the same way, their number of arms, strengths (or amplitudes) $\epsilon_{s,1}$, $\epsilon_{s,2}$ and pattern speeds $\Omega_{s,1}$, $\Omega_{s,2}$ were allowed to differ. Both spiral amplitudes were grown in four solar rotations, as in \S\ref{sec:sp}, after which they were kept constant.

\cite{mq06} adopted a configuration of a primary two-armed and a weaker four-armed spiral wave perturbations, as done previously (e.g., \citealt{amaral97}) . The novelty in this model was the assumption that there was a non-zero relative angular velocity between the two spiral patterns. This introduces an additional parameter - the pattern speed of the secondary spiral wave.

Fig.~\ref{fig:mq06} illustrates how two long-lived spiral density waves propagating with different pattern speeds affect initially circular test-particle orbits. The first column shows a face-on view of particles starting initially on a ring just inside (red) and a ring just outside (green) the CR (dashed black circle) of a two-armed spiral wave of an intermediate strength. Top and bottom plots are identical for the first column. Note the unbroken banana, or ``horseshoe", shape of the orbits. 

The second column of Fig.~\ref{fig:mq06} presents the same initial conditions, but near the 4:1 OLR (top) or the 4:1 ILR (bottom) of a secondary, four-armed wave. This spiral has an amplitude only 1/3 that of the primary, two-armed wave. Only small distortions in the particles' initially circular orbits are apparent. For all three single-spiral simulations no significant time variation was observed.

In the third column of Fig.~\ref{fig:mq06} the two rings of particles are perturbed by both spiral waves shown in each row, shown at t=10 rotation periods, which is about 2.5 Gyr of evolution. Both top and bottom plots are in the reference frame of the corotating, two-armed spiral (not shown). Remarkably, the secondary wave, which has an amplitude of only 1/3 that of the first one, is enough to disrupt the banana orbits near the CR. Moreover, the distances travelled by the particles are now significantly larger, especially in the top panel. The irregular shapes of these orbits suggests the presence of stochasticity, which causes an increase in the particle velocity dispersion, i.e., the disc is heated kinematically with time.  

\begin{figure*}
\centering
\includegraphics[width=\linewidth]{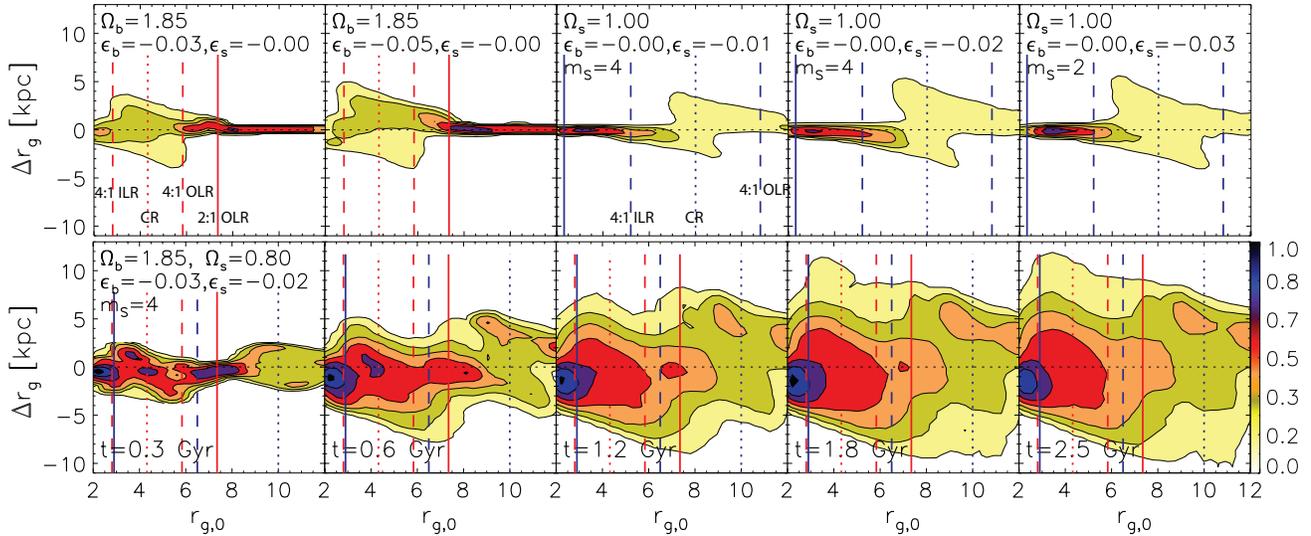}
\caption{
{\bf First row:} Changes in guiding radius (or angular momentum), $\Delta r_g$, as a function of the initial guiding radius, $r_{g,0}$. Bar and spiral amplitudes are indicated in each panel. The first two panels form left to right present simulations with a bar of an intermediate strength ($Q_T=0.25$) and a strong one ($Q_T=0.4$), respectively. The third and fourth panels show a four-armed spiral structure with relative overdensity $\Sigma_s/\Sigma_0=0.2, 0.35$. The rightmost panel presents a simulation with a two-armed spiral structure with an amplitude giving rise to $\Sigma_s/\Sigma_0=0.25$. The different vertical lines depict the 2:1 ILR, 4:1 ILR, CR, 4:1 OLR and 2:1 OLR, as indicated. Note that, depending on the pattern speed, some resonances might not be present in the disc. $|\Delta r_g|$ increases significantly only near the corotation of each perturber. 
{\bf Second row:} Time evolution of the changes in guiding radius, $\Delta r_g$ as a function of initial guiding radius, $r_{g,0}$, for a stellar disc perturbed by both a central bar and spiral structure. The bar and four-armed spiral density wave have amplitudes $\epsilon_b=-0.03$ ($Q_T=0.25$) and $\epsilon_s=-0.02$ ($\Sigma_s/\Sigma_0=0.35$ for m=4). The increase of $\Delta r_g$ with time indicates that stars are being placed on radii different than their birthplaces, i.e., radial migration takes place throughout the disc, even though perturbers are not transient. Figure adapted from \cite{mf10}.
\label{fig:mig}
}
\end{figure*}

The fourth column of Fig.~\ref{fig:mq06} shows the same setup as in the third column but at t=40 rotations, amounting to about 10 Gyr of evolution. At this time the particles appear well mixed, especially at the CR+OLR region (top panel). Concentration of particles at the CR Lagrange points is seen in the bottom panel. 

Previous studies of heating mechanisms have concentrated on heating from transient spiral density waves (e.g., \citealt{carlberg85, desimone04, jenkins92}). However in that case changes in the stellar velocity dispersion only occur during the spirals' growth and decay. \cite{mq06} observed an increase in the velocity dispersion even when the spiral density waves have ceased to grow. Hence, this is a different heating mechanism compared to those explored by these previous investigations.

We note that the above described process causes not only heating but also radial migration, i.e., exchange of angular momentum across the resonance overlap region (discussed in \S\ref{sec:mig}). In other words, some of the particles away from their initial radii in Fig.~\ref{fig:mq06} are there on their apo- or pericenters, while others have shifted their guiding radii altogether.

\cite{mq06} examined how different combinations of pattern speeds, at a given radius, affect the stellar velocity dispersion, finding that this heating mechanism is strongly dependent on the relative velocity between stars and the spiral patterns and thus on the Galactocentric distance. This expected radial variation may be erased when radial mixing of stars is taken into account.  

\subsection{Migration due to the overlap of long-lived patterns}
\label{sec:mig}

As shown by \cite{mq06}, when two perturbers moving at different pattern speeds are imposed on a galactic disc, an increase in the random motions of stars is expected, i.e., the disc heats. In addition, it was found that in regions of resonance overlap stars drift radially with time from their birth radii (see Fig.~\ref{fig:mq06}). This suggests that such a mechanism could be responsible for radial migration in the disc. \cite{mf10} quantified this idea more clearly for the case of a bar and one set of spiral structure, by examining the effect of the simultaneous propagation of the two perturbers. Test-particle simulations were used with a setup similar to that described in \S\ref{sec:heat} above, but where the axisymmetric disc potential from eq.~\ref{eq:axi} is perturbed by a central bar (eq.~\ref{eq:bar}) and a spiral density wave (eq.~\ref{eq:sp}), instead of two spiral waves.

\begin{figure*}
\centering
\includegraphics[width=\linewidth]{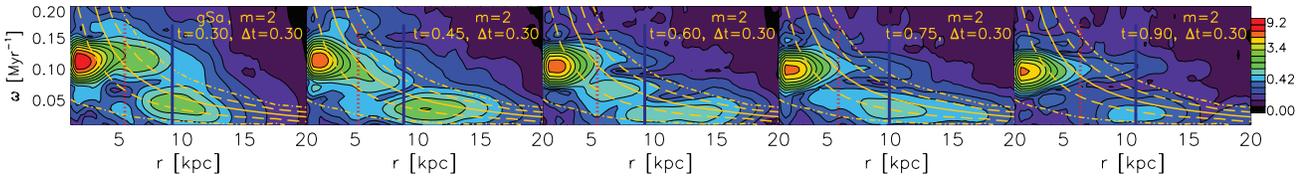}
\caption{
Time evolution of the $m=2$ power spectrum for the gSa model from \cite{minchev12a}. The vertical axis shows the frequency, $\omega=m\Omega$ (in units of km~s$^{-1}$~kpc$^{-1}$). Time outputs are every 150~Myr with a time window $\Delta t=300$~Myr. The orange curves show the resonant loci similarly to Fig.~\ref{fig:res}. The dotted red and solid blue vertical lines indicated the bar CR and 2:1 OLR. It is clear that the outer $m=2$ wave has a stable pattern speed (slowly decreasing as the bar slows down) for over $600~Myr$, while slowly weakening and extending outward with time. In contrast, the inner $m=2$ wave (seen mostly in the middle panel) bounces between the bar and the outer one at the beat frequency of the latter two patterns. Figure adapted from \cite{minchev12a}.
\label{fig:gSa}
}
\end{figure*}

In the first row of Fig.~\ref{fig:mig} we first show how much migration is induced in a galactic disc by a single perturber -- either a bar or a spiral density wave. Plotted are contours of stellar density for the changes in guiding radius, $\Delta r_g$, as a function of the initial guiding radius, $r_{g,0}$. Bar and spiral amplitudes are indicated in each panel. The first two panels form left to right present simulations with a bar of an intermediate strength ($Q_T=0.25$) and a strong one ($Q_T=0.4$), respectively. The third and fourth panels show a four-armed spiral structure with relative overdensity $\Sigma_s/\Sigma_0=0.2, 0.35$. The rightmost panel presents a simulation with a two-armed spiral structure with an amplitude giving rise to $\Sigma_s/\Sigma_0=0.25$. The different vertical lines depict the 2:1 ILR, 4:1 ILR, CR, 4:1 OLR and 2:1 OLR, as indicated. Depending on the pattern speed, some resonances might not be present in the disc. $|\Delta r_g|$ increases significantly only near the corotation of each perturber.

In the second row of Fig.~\ref{fig:mig} we show the time development of the changes of guiding radius in a stellar disc perturbed by both a central bar and spiral structure. Panels from left to right show the temporal evolution of the system for 0.3, 0.6, 1.2, 1.8 and 2.5 Gyr. Both the bar and the four-armed spiral structure have intermediate strengths: $\epsilon_b=-0.03$ ($Q_T=0.25$) and $\epsilon_s=-0.02$ ($\Sigma_s/\Sigma_0=0.35$ for m=4). At the beginning of the simulation $|\Delta r_g|$ increases mainly at the CR of each perturber (dotted lines), which would be the case of linearly adding the individual effects of the bar and spirals seen in the top row. At later times, however, large changes in angular momentum occur, suggesting that non-linear 
effects become important. The changes in guiding radii (angular momentum) increase throughout the simulation timespan manifesting the effect of multiple overlapping patters.

What this experiment demonstrated for the first time was that transient patterns are not a necessity for efficient migration (e.g., \citealt{sellwood02}), but also long-lived, multiple perturbers can be very effective. Because of the hard task of inferring the longevity of spirals in N-body systems, our controlled simulations were the ideal tool for this proof of concept, by allowing us to keep the patterns unevolving.

\subsection{Spiral structure longevity in N-body simulations}
\label{sec:longevity} 

At present the nature of galactic disc spiral structure is not well understood. Though it is generally accepted that spirals are density waves there exist two competing theories:
(i) transient/recurrent spirals, and
(ii) long-lived, steady-state spirals.

Recurrent spiral instabilities have been reported by \cite{sellwood84} and \cite{sellwood89} in their simulations of self gravitating discs. It was argued by \cite{toomre91} that these transient spiral density waves are due to the swing-amplification mechanism as first formulated by \cite{toomre81}.

Alternatively, the concept of quasi-stationary density waves was developed (mostly analitically) by \cite{lin69} and culminated in the work by \cite{bertin89a, bertin89b} and \cite{lowe94}. While thought to always produce short-lived spirals, N-body simulations have been constructed to yield long-lived spiral density waves lasting for over five rotation periods, by introducing an inner Q-barrier to shield the 2:1 ILR \citep{thomasson90, elmegreen93, donner94, zhang96}. 
 
\cite{minchev12a} studied the longevity of spiral structure in N-body Tree SPH simulations from the GalMer database \citep{dimatteo07}. These simulations develop strong bar and spiral structure. The full description and simulation setup can be found in \cite{minchev12a}.

 \cite{minchev12a}'s novel way of estimating the longevity of spirals in N-body discs consisted of following the time evolution of Fourier power spectrograms, which allowed to assessing the development of radial extent, amplitude, and patterns speed. Unlike in most previous works, where power spectrograms are typically computed over time periods of 0.5-1 Gyr, in our analyses we decreased the time window to 300~Myr and computed spectrograms every 150~Myr from $t=150$ to $t=900$~Myr (using outputs every 10~Myr). 

We present the results for the GalMer gSa galaxy model in Fig.~\ref{fig:gSa}. The horizontal and vertical axes show galactic radius, $r$, and angular frequency, $\omega$, respectively; to convert between $\omega$ and the patter speed we use $\Omega=\omega/m$, where $m$ is the multiplicity. The series of panels show the time evolution of the $m=2$ component. At $t=300$~Myr the central bar is seen to extend from the galactic center to $r\approx5.5$~kpc and has a frequency $\omega\approx0.12~{\rm Myr^{-1}}$. A spiral wave of frequency $\omega\sim0.05~{\rm Myr^{-1}}$ is also seen, extending between the bar's CR and the disc break at that time. At later times this wave is always present, increasing in length up to $r\sim18$~kpc at $t=750$~Myr. When animated, this two-armed feature is aways present in the spectrograms, smoothly changing from one time output in the figure to the next. We, therefore, conclude that this pattern has a lifetime $\gtrsim600$~Myr. The smooth decrease in strength, as well as the radial extent with time, is also in agreement with the conclusion that we see the same pattern in all snapshots. In contrast, an inner $m=2$ spiral, seen mostly in the middle panel, exhibits strong variations with time. When animated, this feature appears to be driven by the interaction of the bar with the outer $m=2$ pattern, bouncing back and forth between the two on a timescale consistent with their beat frequency. In other words, every time the (faster) bar encounters the (slower) outer spiral wave, this inner wave is regenerated, speeding up to catch up with the bar ($t=450$ and $t=600$~Myr) and later on slowing down to reconnect with the outer spiral ($t=750$~Myr). Such an inner structure, connecting the bar with the dominant spiral has been reported before and has been proposed to provide an explanation for the nature of the ``long" bar in our Galaxy \citep{athanassoula05, martinez11, romero11, athanassoula12}. 

We estimated above that the lifetime of the two-armed outer spiral wave in our gSa model is $\gtrsim600$~Myr. At its average rotational frequency of $\omega\sim0.04~{\rm Myr^{-1}}$, this corresponds to $\gtrsim4$ rotations, which is relatively long-lived and comparable to estimates found in previous work (e.g., \citealt{thomasson90,elmegreen93,donner94,zhang96}). It should be noted that the model we consider here has a substantial bar, which may be related to the longevity of spirals, especially if mode coupling is present. This spiral lifetime is most likely a low limit because of the quick disc heating taking place in these simulations, due to the lack of gas infall, which, if present, would rejuvenate the spiral structure \citep{bt08}.

\begin{figure*}
\centering
\includegraphics[width=\linewidth]{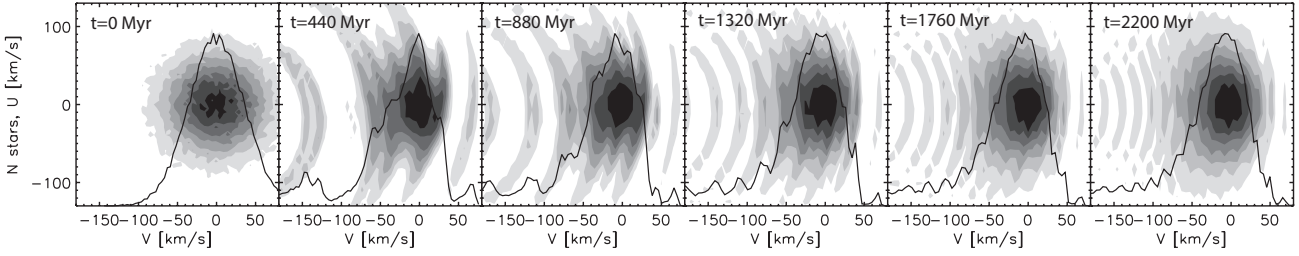}
\caption{
Time development of the simulated solar neighborhood velocity distribution for an axisymmetric disk with initial conditions emulating an impact following a minor merger perturbation. Contours show the $u-v$ plane. The arches seen are energy waves centered on $(u,v+V)=(0,0)$, where $V$ is the LSR velocity. The tangential velocity distribution is shown by the solid line. The sample shown is limited to a radius of 100 pc from our fictitious Sun. As time increases features get closer together as phase wrapping takes place. Figure adapted from \cite{minchev09}.
\label{fig:ringing1}
}
\end{figure*}

\begin{figure}
\centering
\includegraphics[width=\linewidth]{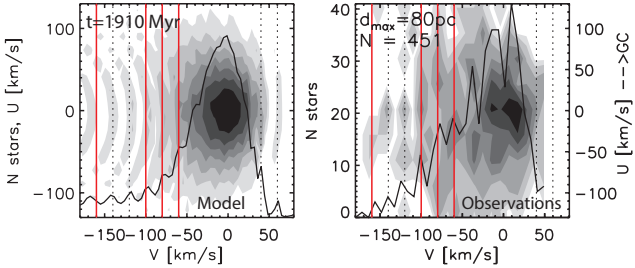}
\caption{ 
Left panel shows the result of the simulation at $t=1910$~Myr after the merger impact. The observed and predicted high-velocity streams are indicated by the solid red and dotted black vertical lines, respectively. A good agreement is found with the data form \cite{nordstrom04} and \cite{schuster06} combined samples in the metallicity range $-1.1<$[Fe/H]$<-0.55$ dex. Figure adapted from \cite{minchev09}.
\label{fig:ringing2}
}
\end{figure}

\section{Constraining recent merger events}

In addition to the moving groups seen in the Hipparcos velocity distribution discussed earlier, high-velocity groups of stars (i.e., $|v|\gtrsim60$~kpc, with $v$ the residual tangential component) have also been identified in various local surveys. Because the bar and spirals have been found to only affect the $u-v$ plane at $|u|,|v|\lesssim50$ km/s. These moving groups are thus usually attributed to accreted populations. For instance, the Arcturus stream \citep{williams08} at $v=-100$ km/s, has been interpreted as originating from the debris of a disrupted satellite \citep{navarro04,helmi06}. Two other groups of stars moving at $v\sim80$ km/s \citep{arifyanto06} and $v\sim-160$ km/s \citep{klement08} were assigned similar origin, based on their kinematics. 

Another interpretation for these high-velocity streams was proposed by \cite{minchev09}, showing that these can be associated with the dynamical effect on the disk by a recent merger event. The distinction here is that these stars are not accreted but belong to the disk and are grouped together in velocity because of the external perturbation. 

How this works is illustrated in Fig.~\ref{fig:ringing1}, which shows the time development of the simulated solar neighborhood velocity distribution for an axisymmetric disk with initial conditions emulating an impact following a minor merger perturbation. Contours show the $u-v$ plane and the solid line is the tangential velocity distribution. 

It is evident that as time increases features get closer together because phase wrapping takes place. There is a particular time when the separation between clumps is similar to that observed, or about 20 km/s. This is shown in Fig.~\ref{fig:ringing2} where the model at $\sim1.9$~Gyr after the perturbation is compared to a combination of the observational samples by \cite{nordstrom04} and \cite{schuster06}.

This model provides an explanation for the following high-velocity streams seen in the solar neighborhood: $v\approx-60$ (HR 1614, \citealt{desilva07})\footnote{Although HR 1614 is established to be a dissolving cluster \citep{desilva07}, it can nevertheless be explained by this model, as long as it is older than the time of merger event.} , $v\approx-80$ \citep{arifyanto06}, $v\approx-100$ (Arcturus), and $v\approx-160$ \citep{klement08}. In addition, it predicts four new features at $v\approx-140, -120, 40$ and 60 km/s.

The separation of the induced velocity streams due to this "disk ringing" could be related to the time of the last merger impact, suggesting an event $\sim1.9$~Gyr ago. The validity of this diagnostic was confirmed with N-body simulations \citep{gomez12a} and found to be consistent with structure in the SEGUE G-dwarf sample \citep{gomez12b}. \cite{gomez13} showed that just such a disturbance from the Sagittarius dwarf galaxy can explain the recent discovery of vertical waves near the Sun in SEGUE \citep{widrow12} and RAVE \citep{williams13}.

\section{Effects of radial migration}

\subsection{Migration and disc thickening}
\label{sec:mig_thick}

Several works have previously suggested that radial migration can give rise to thick disc formation by bringing out high-velocity-dispersion stellar populations from the inner disc and the bulge. Such a scenario was used, for example, in the analytical model of \cite{schonrich09b}, where the authors claimed to explain the Milky Way thick- and thin-disc characteristics (both chemical and kinematical) without the need of mergers or any discrete heating processes. Similarly, the increase of disc thickness with time found in the simulation by \cite{roskar08a} has been attributed to migration in the work by \cite{loebman11}. 

\subsubsection{Migration induces disc flaring in isolated discs}

A first effort to demonstrate how exactly radial migration affects disc thickening in dynamical models was done by \cite{minchev12b}. It was shown that stellar samples arriving from the inner disc have slightly higher velocity dispersions, which will result in them being deposited at higher distances above the galactic midplane. However, the opposite effect arrises from samples arriving from the outer disc (with lower velocity dispersions). Therefore, the {\it overall} migration effect on the disc thickening is minimal throughout most of the disc extend, except in the very inner/outer parts of the disc, where only inward/outward migrators are deposited. This naturally results in disc flaring, as shown in Fig. 7 by \cite{minchev12b}. We explained this as the conservation of vertical action as opposed to conservation of the vertical energy assumed before. Several independent groups, using different simulation techniques and setups, have confirmed that migration does not thicken the disc (\citealt{martig14b, vera-ciro14, grand16, aumer16}). 

\begin{figure}
\centering
\includegraphics[width=\linewidth]{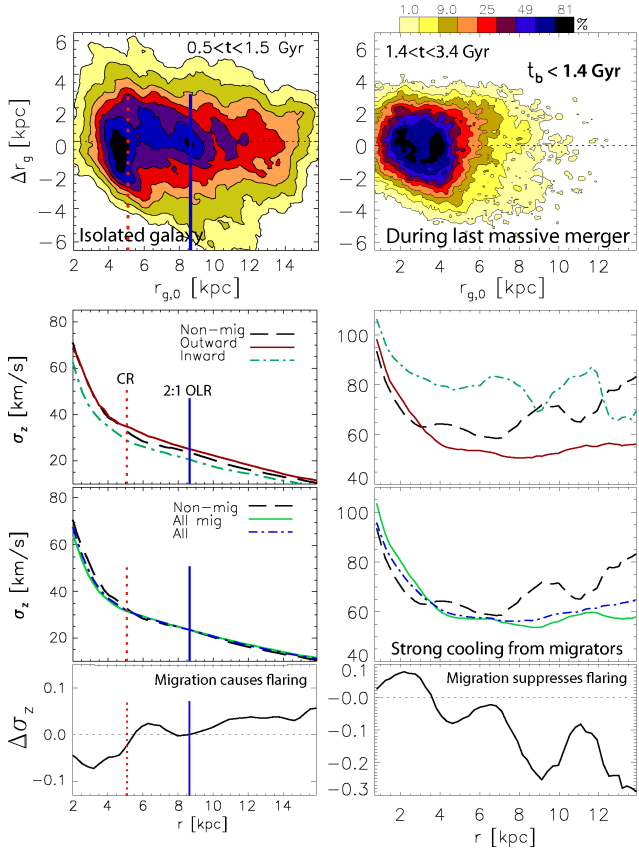}
\caption{
The effect of migration in a quiescent disc evolution (left) and during a massive merger (right). {\bf Left column:} The top panel shows the changes in angular momentum in a time period of 1~Gyr for an isolated N-body disc. The vertical velocity dispersion profiles of inward migrators, outward migrators and the non-migrating population, as indicated, are shown in the second panel. The net effect of migrators can be seen in the third panel. The bottom panel shows the fractional change in velocity dispersion resulting from migration $\Delta\sigma_{\rm z}=(\sigma_{\rm z,all}-\sigma_{\rm z,non\_mig})/\sigma_{\rm z,all}$. Minimal effect from migration is seen on the disc vertical velocity dispersion, except in the innermost and outermost regions, which results in mild disc flaring. 
{\bf Right column:} Same as on the left, but for stars born before the last massive merger event in a galaxy formation model in the cosmological context. In contrast to the isolated disc case, outward migrators cool the outer disc, thus working against disc flaring. Figure adapted from \cite{minchev12b}  and \cite{mcm14}.
\label{fig:disp}
}
\end{figure}

This effect is illustrated in the left column of Fig.~\ref{fig:disp} for an isolated N-body Tree-SPH simulation (the GalMer gSb model) of a barred disc. The top left panel of Fig.~\ref{fig:disp} shows the changes of guiding radius, $\Delta r_g$, vs the initial guiding radius, $r_{g,0}$, during a time period of 1 Gyr; $r_g$ for each star is estimated using the values of the specific angular momentum and rotation curve as $r_g=L/v_\phi$. The percentage of stars in each contour level is given by the color bar on the right. The migration seen in the top panel is solely due to internal (secular) evolution, i.e., the effect of the central bar and spiral arms.

Next we separated migrating from non-migrating stars (in the considered time period) by applying the technique described by \cite{minchev12b}. This consists of separating stars in a given radial bin into migrators and non-migrators as follows. Non-migrators are those particles found in the selected radial bin at both the initial and final times, while migrators are those that were not present in the bin initially but are there at the final time. We distinguish between outward and inward migrators -- those initially found at radii smaller or larger than the annulus considered, respectively. This is done for radial bins over sampling the entire radial extent of the disc. See \cite{minchev12b} for more details.

In the second top-to-bottom left panel of Fig.~\ref{fig:disp} we plot the vertical velocity dispersion profiles of inward migrators, outward migrators and the non-migrating population, as indicated. We find that stars arriving from the inner disc are slightly hotter than the non-migrating populations but the ones arriving from the outer disc are cooler. In the third top-to-bottom left panel it can be seen that the overall effect of all migrators is negligible to the overall vertical velocity dispersion and, thus, to the disc thickening. Some cooling inside $r=3$ kpc is seen, which is related to the accumulation of inward migrating stars with small vertical actions in that region.

Finally, to quantify the changes to the disc vertical velocity dispersion resulting from migration in the given period of time, we plot the fractional changes in the bottom left panel of Fig.~\ref{fig:disp}. As in \cite{minchev12b},  we estimate these as 
\begin{equation}
\label{eq:1}
\Delta\sigma_{\rm z}=(\sigma_{\rm z,all}-\sigma_{\rm z,non\_mig})/\sigma_{\rm z,all}, 
\end{equation}
where $\sigma_{\rm z,all}$ and $\sigma_{\rm z,non\_mig}$ are the vertical velocity dispersions for the total population and the non-migrators, respectively. Fluctuations of less than 10\% are seen around $\Delta \sigma_z=0$. The positive slope of $\Delta \sigma_z$ with radius indicates an overall cooling/heating in the galactic center/disc outskirts. This is what gives rise to the flaring described by \cite{minchev12b}. The degree of flaring cause by migration is much less than that by infalling satellites (40\% vs 800\%, respectively, over $\sim4$ scale-lengths; compare \citealt{minchev12b} to \citealt{bournaud09}).

\subsubsection{Migration suppresses disc flaring when infalling satellites are present}

It is well known from both observations and cosmological simulations that minor mergers take place in the formation of galactic discs. The intensity of these interactions with the host disc decrease with redshift but can persist until today, as evident in the Milky Way (e.g., the Sagittarius dwarf galaxy, \citealt{ibata94,ibata95}). Such interactions will have the effect of heating more the disc outskirts at any given time of the disc growth (more so at high redshift), because of the low mass density there. 

Interestingly, and to complicate matters, when this more realistic scenario is considered, migration has the opposite effect on the disc vertical profile compared to the effect of an isolated galaxy -- the role of outward and inward migrators is reversed in that they now cool and heat the disc, respectively. 
We next demonstrate this in the right column of Fig.~\ref{fig:disp}, which presents a simulation by \cite{martig12} used for the chemo-dynamical model presented by (\citealt{mcm13,mcm14}; see \S\ref{sec:chem}). We consider all stars born right before the last massive merger encounters the disc at $t=1.4$~Gyr. The strong redistribution of angular momentum seen in the $r_{g,0}-\Delta r_g$ plane is caused both by the tidal effect of the satellite, which plunges through the galactic center, and the strong spiral structure induced in the gaseous component. 

Examining the right second top-to-bottom panel, it is remarkable that inward and outward migrators {\it during the merger} have reversed roles compared to the case of an isolated disc (left column), where stars migrating inwards have positive contribution to $\sigma_z$ and those migrating outward cool the disc. 

The net effect of migrators during the merger can be seen in the third top-to-bottom right panel of Fig.~\ref{fig:disp}. The overall contribution to the vertical velocity dispersion from the migrating stars during the merger is negative, in the sense that it is lower than that of the stars which did not migrate. We emphasize that we only considered stars born before the merger took pace, therefore, the effect seen is not related to the accreted population. 

In the bottom right panel of Fig.~\ref{fig:disp} we can see a negative slope for $\Delta \sigma_z$ vs $r$, in contrast to the left panel, indicating that the disc flaring induced by the merging satellite is being counteracted by outward migrating stars. A decrease in $\sigma_z$ of up to 30\% in the disc outskirts is seen. 

Fig.~\ref{fig:disp} showed that during a massive merger sinking deep into the disc center, migrators cool the outer disc, thus working against disc flaring. This is related to the stronger effect of mergers on the outer disc, owing to the exponential decrease in the disc surface density. Stars arriving from the inner parts during a merger, would therefore be cooler than the rest of the population. Note that flaring for a given mono-age population, nevertheless, still results from infalling satellites and that gives rise to the formation of a thick disc, as discussed next.

\section{Formation of galactic thick discs by the flaring of mono-age populations}
\label{sec:thick}

Stellar disc density decomposition into thinner and thicker components in external edge-on galaxies find that thicker disc components have larger scale-lengths than the thin discs (e.g., \citealt{yoachim06, pohlen07, comeron12}). While this is consistent with results for the Milky Way when similar morphologically (or structural) definition for the thick disc is used (e.g., \citealt{robin96, ojha01, juric08}), it is in contradiction with the more centrally concentrated older or [$\alpha$/Fe]-enhanced stellar populations (e.g., \citealt{bensby11, cheng12b, bovy12a}). This apparent discrepancy may be related to the different definition of thick discs - morphological decomposition or separation in chemistry.

Additionally, while no flaring is observed in external edge-on discs \citep{vanderkruit82, degrijs98, comeron11}, numerical simulations suggest that flaring cannot be avoided due to a range of different dynamical effects. The largest source is most likely satellite-disc interactions (e.g., \citealt{bournaud09, villalobos08}), which have been found to increase an initially constant scale-height by up to a factor of $\sim10$ in 3-4 disc scale-lengths. We showed in the left row of Fig.~\ref{fig:disp} that purely secular evolution (in the absence of external perturbations) also causes flared discs, due to the redistribution of disc angular momentum. Other sources of disc flaring include misaligned gas infall \citep{scannapieco09} and reorientation of the disc rotation axis \citep{aumer13a}. 

\begin{figure}
\centering
\includegraphics[width=7 cm]{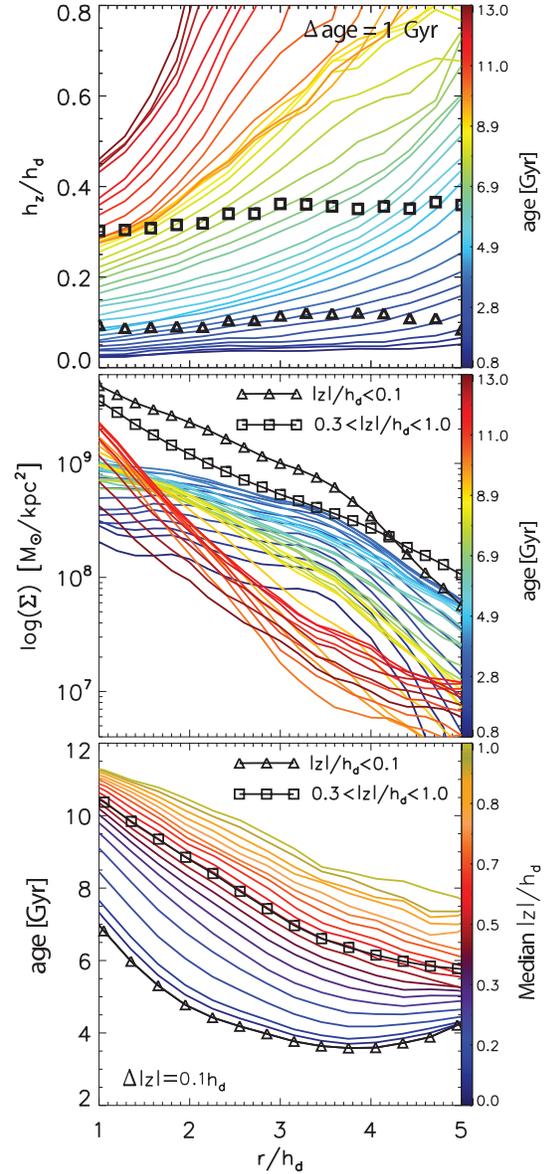}
\caption{
{\bf Top:} Variation of disc scale-height, $h_z$, with galactic radius for a cosmological disc formation simulation. Color lines show mono-age populations, as indicated. Overlapping bins of width $\Delta$age $=1$~Gyr are used. Overlaid also are the thin (triangles) and thick (squares) discs obtained by fitting a sum of two exponentials to stars of all ages. No significant flaring is found for the thin and thick discs. 
{\bf Middle:} Disc surface density radial profiles of mono-age populations. Older discs are more centrally concentrated, which explains why flaring diminishes in the total population. Also shown are the surface density profiles of stars close to (triangles) and high above (squares) the disc midplane. The thicker disc component extends farther out than the thin one, consistent with observations of external galaxies.
{\bf Bottom:} Variation of mean age with radius for samples at different distance from the disc midplane, as indicated. Slices in $|z|$ have thickness $\Delta|z|=0.1h_d$. Overlaid are also the age radial profiles of stars close to (triangles) and high above (squares) the disc midplane. Age gradients are predicted for both the (morphologically defined) thin and thick discs. Adapted from \cite{minchev15}.
\label{fig:thick1}
}
\end{figure}

\begin{figure}
\centering
\includegraphics[width=\linewidth]{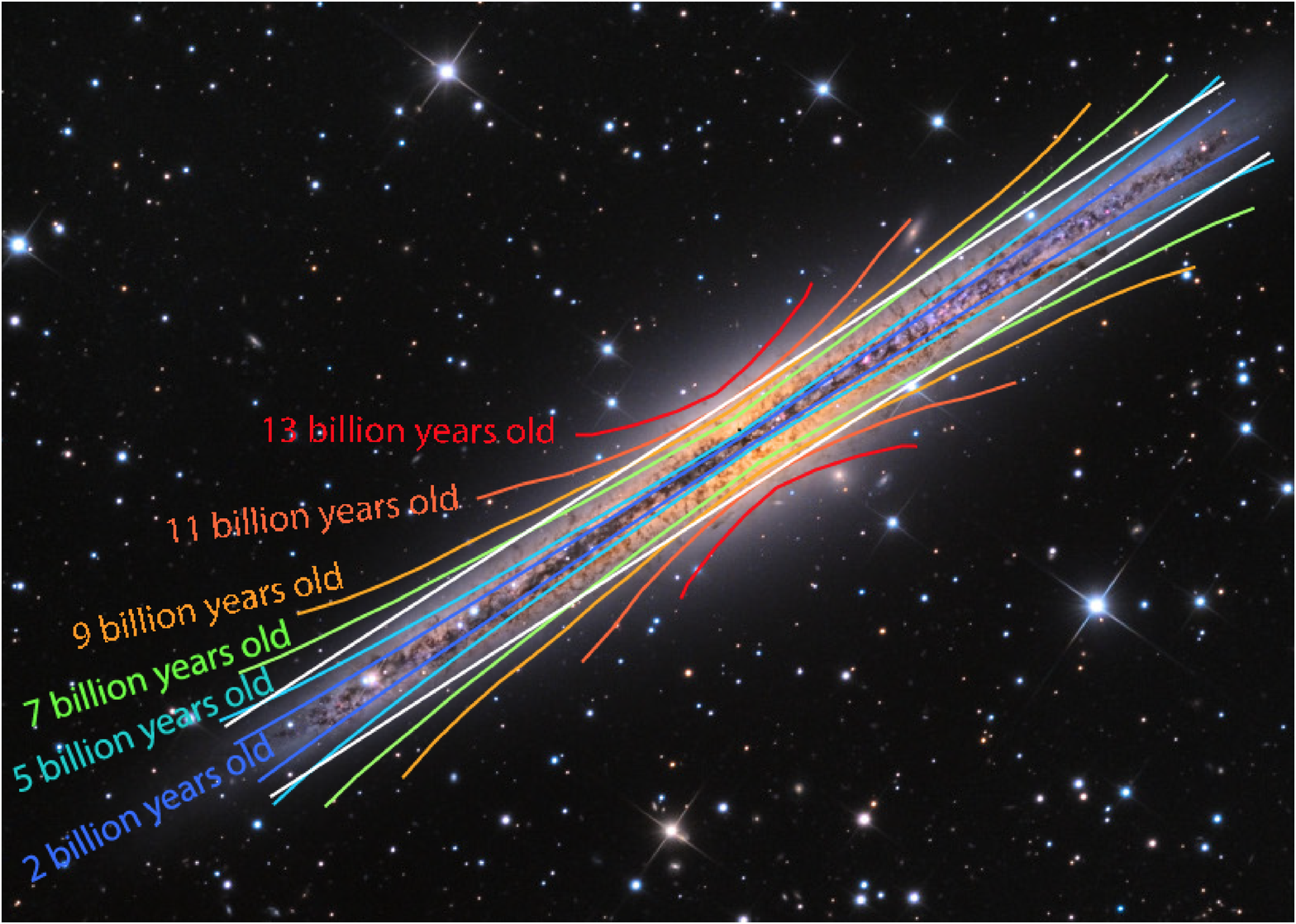}
\caption{
Background image is of the Milky Way analog galaxy NGC 891 (Credit: Adam Block, Mt. Lemmon SkyCenter, University of Arizona). Overlaid are color curves that show the flares from groups of stars with similar ages, as in top panel of Fig.~\ref{fig:thick1} but scaled to the observed galaxy. When all stars are put together, the disc has roughly constant thickness, illustrated by the straight white lines. 
\label{fig:thick2}
}
\end{figure}

\cite{minchev15} studied the formation of thick discs using two suites of simulations of galactic disc formation, one of which we present here. This is a full cosmological zoom-in hydro simulation, using initial conditions from one of the Aquarius Project haloes \citep{springel08, scannapieco09}. The technique used here is Tree-PM SPH with 300~pc spatial and $4.4\times10^5$~M$_{\odot}$ mass resolution. This is a non-barred galaxy with stellar mass $5.5\times10^{10} M_{\odot}$ and a disc scale-length $h_d=4$~kpc. Further details about this simulation can be found in \cite{aumer13b}, their model Aq-D-5.

The simulation forms an initial central component during an early epoch of violent merger activity. Gas-rich mergers supply the initial reservoir of gas at high redshift and merger activity decreases with redshift, similarly to what is expected for the Milky Way. This inside-out disc formation results in a centrally concentrated older stellar population. The general formation and evolutionary behavior of this model is similar to many recent simulations in the cosmological context (e.g., \citealt{brook12, stinson13}). 

We decomposed the stellar discs into mono-age populations, i.e., narrow bins of age, where we used $\Delta$age=1~Gyr. It was found that single exponentials provided good fits to the column density in the vertical direction for all age bins and at all radii, in agreement with \cite{martig14a}. In contrast, to properly fit the vertical density of the total stellar population required the sum of two exponentials.

In the top panel of Fig.~\ref{fig:thick1} we plot the scale-height variation with galactocentric radius, $r$, in the region 1-5 disc scale-lengths, $h_d$. Both the radius and scale-height, $h_z$, are in units of $h_d$. It can be seen that significant flaring is present, which increases for older coeval populations. 

In contrast to the flaring found for all but the youngest mono-age populations, the thin and thick disc decomposition of the total stellar population including all ages, results in no apparent flaring. This is shown by the triangle and square symbols overlaid in the top panel of Fig.~\ref{fig:thick1}.

What is the reason for the flaring of mono-age discs? In numerical simulations flaring is expected to result from a number of mechanisms related to galactic evolution in a cosmological context (e.g., \citealt{bournaud09, kazantzidis08, villalobos08, aumer13a}). Even in the absence of environmental effects, flaring is unavoidable due to secular evolution alone (radial migration caused by spirals and/or a central bar, \citealt{minchev12b}). It should be stressed here that, while migration flares discs in the lack of external perturbations, during satellite-disc interactions it works {\it against} disc flaring (\citealt{mcm14}, see \S\ref{sec:mig_thick}). Yet, this is not sufficient to completely suppress the flaring induced by orbiting satellites, as evident from the top row of Fig.~\ref{fig:thick1}. This suggests that external effects are much more important for the disc flaring in this simulation. Because the mass and intensity of orbiting satellites generally decreases with decreasing redshift, so does the flaring induced. It can be expected that at a certain time secular evolution takes over the effect of external perturbations.\footnote{\cite{minchev14} suggested that the time at which internal evolution takes over can also be inferred from the shape of the [$\alpha$/Fe]-velocity dispersion relation of narrow metallicity samples.}

What is the reason for the lack of flaring in the total disc population? In an inside-out formation scenario, the outer disc edge, where flaring is induced, moves progressively from smaller to larger radii because of the continuous formation of new stars in disc subpopulations of increasing scale-length. At the same time the frequency and masses of perturbing satellites decreases. Because of the inside-out disc growth, which results in more centrally concentrated older samples (see Fig.~\ref{fig:thick1}, middle panel), the younger the stellar population, the further out it dominates in terms of stellar mass. The geometrically defined thick disc, therefore, results from the imbedded flares of different coeval populations, as seen in the top panel of Fig.~\ref{fig:thick1}. 

The bottom panel of Fig.~\ref{fig:thick1} shows that a geometrical thick disc is expected to have a negative age gradient. For this particular simulation the mean age decreases from $\sim10.5$ to $\sim6$~Gyr in four disc scale-lengths. Such an age drop of mean stellar age at high distances, $|z|$, from the disc midplane explains the inversion in $[\alpha$/Fe] gradients with increasing mean $z$ found by \cite{anders14} in APOGEE data.

\begin{figure*}
\centering
\includegraphics[width=\linewidth]{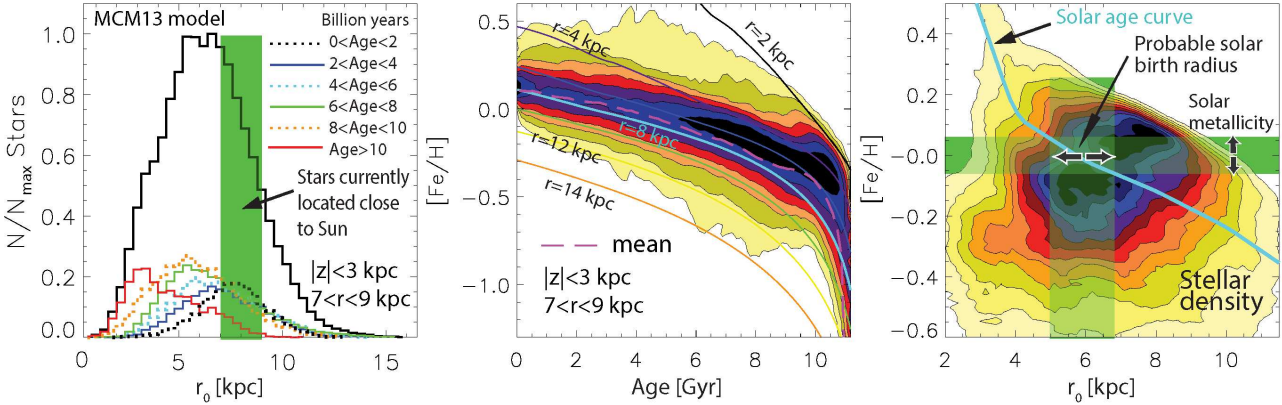}
\caption{
{\bf Left}: Illustration of the effects of radial migration - birth radii of stars ending up in the simulated solar vicinity (green shaded strip) at the final simulation time (MCM13 model). The solid black curve shows birth radii of all local stars, while the color-coded curves show distributions for six different age groups. 
{\bf Middle}: The resulting age-metallicity relation for stars in located at $7<r<9$~kcp at the final time (as in the left panel). The chemical evolution of different radial bins are overlaid. Despite the scatter, the mean relation (pink dashed curve) remains close to the local evolution curve.
{\bf Right}: Search for the solar birth radius - the intersection of solar metallicity (shaded horizontal strip) with solar age (cyan curve) results in a range of possible solar birth radii (vertical shaded strip). Figure adapted from MCM13.
\label{fig:mcm13}
}
\end{figure*}

\section{Chemo-dynamical modeling of the Milky Way}

So far we have focused on the dynamics of the Milky Way disc, which tells us mostly about its current state. To be able to go back in time and infer the Milky Way evolutionary history, however, we need to include both stellar chemical and age information.

\subsection{Summary of Milky Way chemo-dynamical evolution modeling techniques}

A major consideration in a disc chemo-dynamical model is taking into account the effect of radial migration, i.e., the fact that stars end up away from their birth places. Below we briefly summarize models which include radial migration. 
\newline$\bullet$ Semi-analytical models tuned to fit the local metallicity distribution, velocity dispersion, and chemical gradients, etc., today (e.g., \citealt{schonrich09a, kubryk15}) or Extended distribution functions \citep{sanders15}: 
\newline$-$ Easy to vary parameters
\newline$-$ Provide good description of the disc chemo-kinematic state today
\newline$-$ Typically not concerned with the Milky Way past history
\newline$-$ Time and spatial variations of migration efficiency due to dynamics resulting from non-axisymmetric disc structure is hard to take into account.
\newline$\bullet$ Fully self-consistent cosmological simulations (e.g., Kawata and Gibson 2003; Scannapieco et al. 2005; Kobayashi and Nakasato 2011; Brook et al. 2012):
\newline$-$ Dynamics self-consistent in a cosmological context
\newline$-$ Can learn about disc formation and evolution
\newline$-$ Not much control over final chemo-kinematic state
\newline$-$ Problems with SFH and chemical enrichment due to unknown subgrid physics
\newline$-$ Much larger computational times needed if chemical enrichment included.
\newline$\bullet$ Hybrid technique using simulation in a cosmological context + a classical (semi-analytical) chemical evolution model (\citealt{mcm13}, hereafter MCM13):
\newline$-$ Avoids problems with SFH and chemical enrichment in fully self-consistent models
\newline$-$ Can learn about disc formation and evolution
\newline$-$ Not easy to get Milky Way-like final states.

Below we focus on the results of the latter model.

\section{The MCM13 hybrid chemo-dynamical model}
\label{sec:chem}

To properly model the Milky Way it is crucial to be consistent with observational constraints at redshift $z=0$, for example, a flat rotation curve, a small bulge, a central bar of an intermediate size, gas to total disc mass ratio of $\sim0.14$ at the solar vicinity, and local disc velocity dispersions close to the observed ones.

It is clear that cosmological simulations would be the natural framework for a state-of-the-art chemo-dynamical study of the Milky Way. Unfortunately, as discussed by MCM13, a number of star formation and chemical enrichment problems still exist in fully self-consistent simulations. We have, therefore, resorted to the next best thing -- a high-resolution simulation in the cosmological context coupled with a pure chemical evolution model.

The simulation used is part of a suite of numerical experiments presented by \cite{martig12}, where the authors studied the evolution of 33 simulated galaxies from $z=5$ to $z=0$ using the zoom-in technique described by \cite{martig09}. This technique consists of extracting merger and accretion histories for a given halo in a $\Lambda$-CDM cosmological simulation and then re-simulating at much higher resolution (150~pc spatial, and 10$^{4-5}$~M$_{\odot}$ mass resolution).

Originally, our galaxy has a rotational velocity at the solar radius of 210 km/s and a scale-length of $\sim5$~kpc. To match the Milky Way in terms of dynamics, at the end of the simulation we downscale the disc radius by a factor of 1.67 and adjust the rotational velocity at the solar radius to be 220 km/s, which affects the mass of each particle according to the relation $GM\sim v^2r$, where $G$ is the gravitational constant. This places the bar's CR and 2:1 OLR at $\sim4.7$ and $\sim7.5$~kpc, respectively, consistent with a number of studies (e.g., \citealt{dehnen00, mnq07, minchev10}; see \S\ref{sec:bar}). At the same time the disc scale-length, measured from particles of all ages in the range $3<r<15$~kpc, becomes $\sim3$~kpc, in close agreement with expectations in the Milky Way. 

\subsection{Birth radius distribution of solar vicinity stars}

To illustrate the migration efficiency in the MCM13 model, in the left panel of Fig.~\ref{fig:mcm13} we show the birth radii of stars ending up in a solar neighborhood-like location ($7<r<9$~kpc, green shaded strip, and $|z|<3$~kpc) after 11.2~Gyr of evolution. The solid black line shows the total population, which peaks close to $r_0=6$~kpc due to radial migration. The entire sample is also divided into six age-groups, shown by the curves of different colors and line-styles. The strongest effect from radial migration is found for the oldest stars (red curve), whose distribution has a maximum at $r\approx3$~kpc, or inside the bar's CR (at $\sim4.7$~kpc). Note that locally born stars of all ages can be found in the solar neighborhood. A relatively smooth transition of the peak, from older to younger groups of stars, is observed; this is expected, since even for a constant migration efficiency older stars would be exposed longer to perturbations. While a wide range of birth radii is seen for all age groups, the majority of youngest stars were born at, or close to the solar neighborhood bin. 

\begin{figure*}
\includegraphics[width=\linewidth]{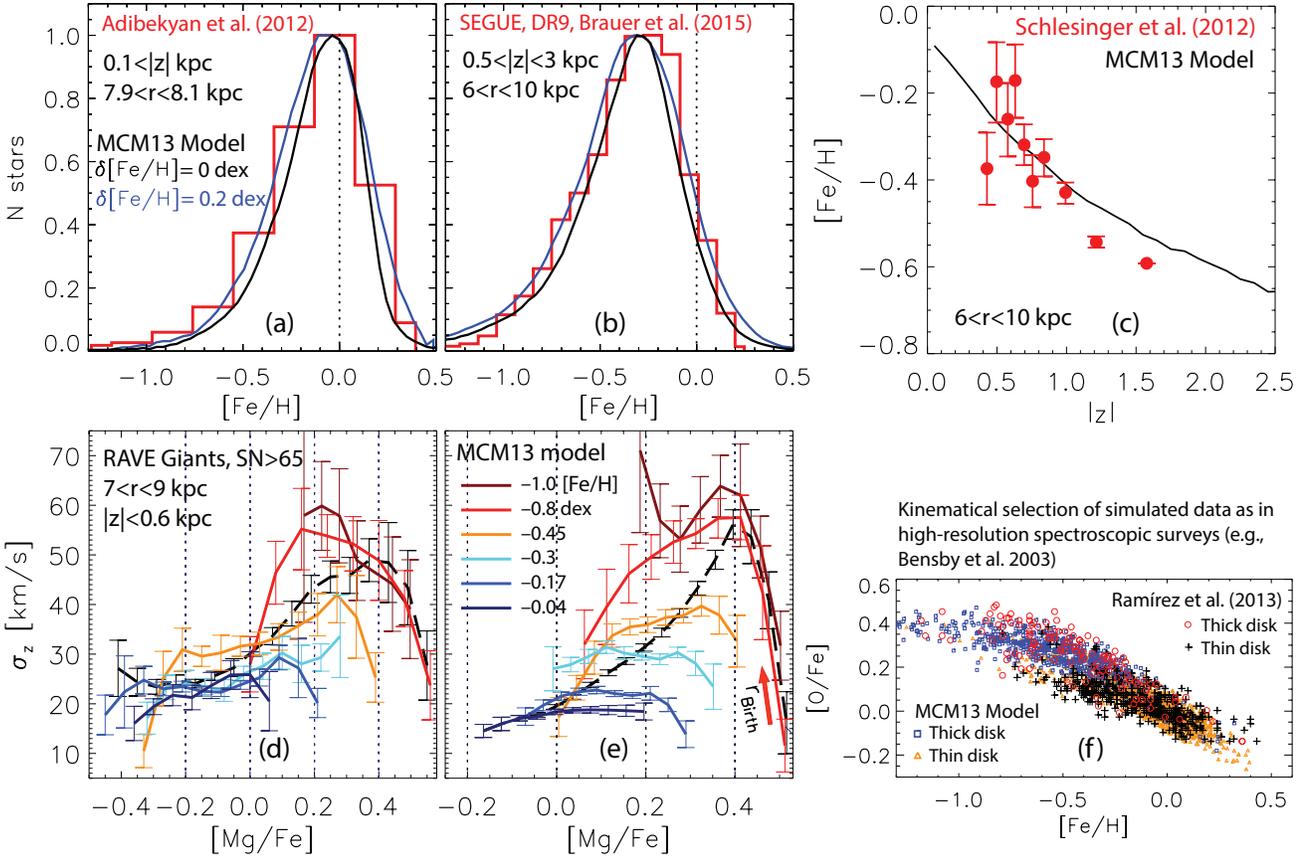}
\caption{
Comparison between the prediction of the MCM13 model and observations. 
{\bf Panels (a) and (b):} The red histograms show data from Adibekyan et al. (2012) and Bauer et al. (in preparation); the black and blue curves show the model with and without convolved error, respectively. The metallicity peak shifts to lower [Fe/H] for both the data and model, when distance from the disc plane increases. 
{\bf Panel (c):} Metallicity variation with distance from the disc plane for SEGUE G-dwarf data (red). 
{\bf Panels (d) and (e):} Variations of vertical velocity dispersion with [Mg/Fe] for RAVE giants (d) and for the model (e). 
{\bf Panel (f):} Comparison to the high-resolution data by Ramirez et al. (2013) using the kinematical selection of \cite{bensby03}. A shift in the model [O/Fe] of 0.05 dex (within the uncertainty) has been applied. 
Panels (a), (b), (c), (f) are from MCM13 and panels (d) and (e) are from Minchev et al. (2014a).}
\label{fig:chem}      
\end{figure*}

\subsection{The age-metallicity relation}

In the middle panel of Fig.~\ref{fig:mcm13} shows stellar density contours of the resulting age-metallicity relation for the ``solar" cylinder shown in the left panel. Overlaid on top of the contours is our input chemical evolution model for different initial radii, as indicated, giving an insight into the origin of stars found in this localized region. The excess of stars above the local curve (cyan) is due to migrators coming from the inner disc. Similarly, contours below the local curve result from stars originating in the outer disc. 

The mean metallicity, shown by the pink dashed curve, is found to follow closely the in-situ born population (cyan curve). Some flattening is observed, mostly for ages $\gtrsim$9~Gyr, however the final distribution is by no means flat. The reason for this minor effect on the local metallicity gradient, despite the strong migration, is the fact that at the Sun's intermediate distance from the Galactic center the change in metallicity arising from stars migrating from the inner regions is mostly compensated for by stars arriving from the outer disc.

\subsection{The birth place of the Sun}

With the care taken into defining a proper solar radius in a simulation with Milky Way characteristics we can made an estimate for the solar birth radius. The right panel of Fig.~\ref{fig:mcm13} displays density contours of the $r_0$-[Fe/H] plane for all local stars. The cyan curve shows the input solar-age metallicity gradient. This is taken to be 4.6~Gyr look-back time, consistent with, e.g., \citealt{bonanno02, christensen09, houdek11}. Assuming an error of $\pm0.06$~dex in [Fe/H], we find a possible Sun birth radius in the range $5.0<r_{\odot, birth}<6.8$~kpc (where the horizontal green, transparent strip meets the cyan curve)\footnote{This estimate is also dependent on the solar metallicity error assumed, e.g., MCM13 used $\rm [Fe/H]_{err}\pm0.1$~dex obtaining $4.4<r_{\odot, birth}<7.7$~kpc.}. This result is dependent on the migration efficiency in the simulation and is in good agreement with the estimate of $6.6\pm0.9$~kpc by \cite{wielen96}.

\subsection{Constraining the entire Milky Way merger history}

Using data from the RAVE and SEGUE surveys, \cite{minchev14} presented a previously unknown chemo-kinematic relation, where the variation of stellar velocity dispersion with [Mg/Fe] ratios is reversed beyond 0.4 dex (Fig.~\ref{fig:mcm13}, panel d). This is unexpected because stars with higher [Mg/Fe] ratios, at a given radius, should be older and, thus, should possess larger random energies. An excellent match by the MCM13 chemo-dynamical model (Fig.~\ref{fig:mcm13}, panel e) revealed that this new relation offers a unique way to recover the Milky Way merger history, where the peak of each metallicity subpopulation gives away the time and mass of the perturber and the decline on the right is related to older cooler stars migrating from the inner disk. This observational relation has been also found in Gaia-ESO data \citep{guiglion15} and is currently being investigated with APOGEE data.

By using our range of models, we will quantify global structures in the Milky Way disk caused by mergers as a function of satellite mass, inclination, and compactness. We may discover, for example, that bar formation is related to a previous merger, tidal deformation, or gas accretion event. We may also find evidence for kinematic and metallicity variations in the disk directly related to episodes of gas accretion tied to merger events.

\begin{figure}
\centering
\includegraphics[width=\linewidth]{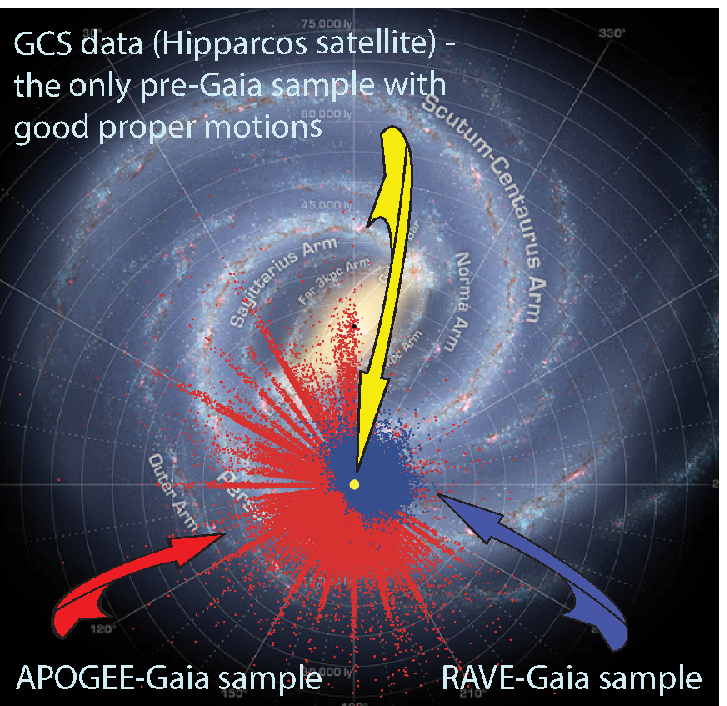}
\caption{
Illustrating the vast increase of stars with precise 6D kinematical information and chemical abundances after Gaia's first two data releases. Distribution of RAVE-Gaia and APOGEE-Gaia stellar samples overlaid on top of R. Hurt's map of the Milky Way (SSC-Caltech). The Galactic disc rotation is in the clockwise direction. 
\label{fig:mw}
}
\end{figure}

\subsection{Comparison with observations}

The MCM13 model has been found to match simultaneously the following observational constraints, some of which are reproduced in Fig.~\ref{fig:chem}:
\newline$\bullet$ The disc morphology, scale-length and rotation curve today (see MCM13)
\newline$\bullet$ The local age-velocity dispersion relation (Sharma et al. 2014)
\newline$\bullet$ The more centrally concentrated [$\alpha$/Fe]-enhanced (old) disc (Bensby et al. 2011; Bovy et al. 2012; see Minchev, Chiappini, and Martig 2014b, Fig.11)
\newline$\bullet$ The distribution of scale-heights for mono-abundance subpopulations found in SEGUE G-dwarfs (MCM13, Fig.13)
\newline$\bullet$ The reversal of the radial [$\alpha$/Fe] and metallicity gradients (e.g., APOGEE - Anders et al. 2014), when sample distance from the disc mid-plane is increased (Minchev et al. 2014b)
\newline$\bullet$ The MDF for stellar samples at different distance from the disc midplane (Fig.~\ref{fig:chem}, a, b)
\newline$\bullet$ The metallicity variation with vertical distance from the plane (Fig.~\ref{fig:chem}, c)
\newline$\bullet$ The inversion in velocity dispersion relation in RAVE and SEGUE (Fig.~\ref{fig:chem}, d, e; Minchev et al. 2014; Guiglion et al. 2015)
\newline$\bullet$ The [Fe/H]-[$\alpha$/Fe] plane (Fig.~\ref{fig:chem}, f)
\newline$\bullet$ The age-[$\alpha$/Fe] relation (Fig.~2 by \cite{minchev16a})
\newline$\bullet$ The age-[Fe/H] relation (Fig.~2 by \cite{minchev16a})
\newline$\bullet$ The flaring of mono-abundance populations (assuming similarity to mono-age populations) found by Bovy et al. (2015) and predicted earlier by Minchev et al. (2015), where Model 1 in the latter paper presents the same galaxy as the one used for the MCM13 chemodynamical model.

It is impressive that this simple match between a simulation in the cosmological context with Milky Way characteristics and a classical Milky Way chemical evolution model, where for both the SFH is such that the disc grows inside-out, was able to account for so many chemo-kinematic relations. This model needs to be improved to account also for the gap in the [$\alpha$/Fe]-[Fe/H] relation seen for a long time in high-resolution spectroscopic data in the solar vicinity (e.g., \citealt{fuhrmann04, bensby13, adibekyan13}) and more recently found in a range of Galactic radii using APOGEE data \citep{anders14, nidever14}. This separation into low- and high-[$\alpha$/Fe] sequences (sometime referred to as the thin and thick discs)\footnote{Note that this is the chemical definition of thin and thick disks, which is different than a definition using morphology (or geometry). See discussion in \S\ref{sec:thick}.} is most likely the result of a gap in the SFH at high redshift, as first suggested in the Two-Infall Model by \cite{chiappini97}.

\section{Discussion and conclusions}

In this work we reviewed some methods used to constrain the dynamics of the Milky Way disc, in particular, its central bar, spiral structure, and the formation of the thick disc. While kinematics alone can be used to constrain the current dynamical state of the disc, chemical information and accurate age estimates are needed to understanding the Milky Way formation, with the help of detailed chemo-dynamical evolution modeling.

We emphasize the importance of taking into account today's Galactic disc morphology in the construction of chemo-dynamical models, which ensures that resonances are placed at the appropriate Galactic radii. Although there is no guarantee that a model displaying similar final state to that of the Milky Way had the same evolutionary history, taking as a constraint the only phase-space snapshot we have of the Galaxy is a must. 

The field of Galactic Archaeology will soon be transformed by the Gaia mission, which will provide precise positions, proper motions, radial velocities, and stellar parameters for up to 1 billion stars of the Milky Way and will map distances out to 15 kpc from the Sun. This can be contrasted to a total of a few million Milky Way stars observed to date, only a fraction of which lie farther than ~2-3 kpc from the Sun. Stars with accurate 6D kinematics are currently available only within ~0.1 kpc from the Sun (yellow dot in Fig.~\ref{fig:mw}) provided by Hipparcos and GCS. This will soon change already with the first and second Gala data releases expected in 2016 and 2017, respectively, which will complement existing spectroscopic data (e.g., RAVE-Gaia and APOGEE-Gaia synergies led by the Leibniz-Institut f\"{ur} Astrophysik Potsdam AIP) and will dramatically increase the disc coverage of stars with precise kinematics (see Fig.~\ref{fig:mw}).

Availability of accurate ages is also very important to make progress in the field of Galactic Archaeology by breaking degeneracies in chemo-dynamical models. This has recently become evident with the unexpected results of \cite{chiappini15} and \cite{martig15}, who used CoRoT \citep{baglin06} and Kepler \citep{gilliland10} asteroseismic ages, respectively, combined with APOGEE chemical information, to show the existence of significantly young high-[$\alpha$/Fe] stars (but see also \citealt{yong16}). This is a largely unexpected result for chemical evolution modeling, because [$\alpha$/Fe] has been thought to always be a good proxy for age. Stellar ages for much larger samples and broader disc coverage are expected in the very near future from Kepler-2 and the Gaia mission. These will help break degeneracies and refine chemo-dynamical models, thus bringing us a step closer to understanding the formation of our Galaxy.

\bibliographystyle{an}
%


\end{document}